\documentclass[11pt]{article}
\usepackage{psfig}
\usepackage{latexsym}
\newcommand{\Grad}{\mbox{grad}}
\newcommand{\Div}{\mbox{div}}

\begin{document}
\renewcommand{\theequation}{\arabic{equation}}
\renewcommand{\thefigure}{\arabic{figure}}
\renewcommand{\thetable}{\arabic{table}}
\newenvironment{eqn}{\begin{equation}}{\end{equation}}
\newcommand{\beq}{\begin{eqn}}
\newcommand{\eeq}{\end{eqn}}
\vspace*{30 true mm}
\begin{raggedright}
{\Large \bf Modellization of hydraulic fracturing of porous 
materials\footnote{Reference: Advances in Fracture Mechanics, 
{\em Fracture of Rock} ed. M.~H.~Alibadi, (WIT Press, Southhampton, 1999),
p.227-61.}
} \\
\large
{F. Tzschichholz$^{1}$ \& M. Wangen$^2$}\\
{\em $^1$ Department of Physics, Norwegian University of Science \\
and Technology, N--7034 Trondheim, Norway \\
EMail: frank@ica1.uni-stuttgart.de}\\
{\em $^2$ Institute for Energy Technology, N--2007 Kjeller, Norway \\
EMail: magnus@ife.no}\\
\end{raggedright}
\section*{Abstract}
\small
\setlength{\baselineskip}{12pt}
We review microstructural fracture growth models suitable for the study 
of hydraulic fracture processes in disordered porous materials and
present some basic results.
It is shown that microstructural models exhibit certain 
similarities to corresponding theories of
continua. These similarities are most easily demonstrated for simple 
crack geometries, i.e., straight cracks (finite size scalings). 
However, there exist even scaling relations which are
completely independent of the particular employed crack structure.
Furthermore it is demonstrated that disorder in cohesional/flow
properties can influence the crack growth and the resulting fracture 
geometry in an essential way.
\setlength{\baselineskip}{12pt}
\section{Introduction}
\label{sec:intro}
\normalsize
The classification and prediction of breaking processes within solid
materials forms an extremely difficult problem in engineering 
science as well as in solid state physics.  
This is even true for brittle materials exhibiting the simplest 
rheological behavior, i.e. no remanent deformations.
Continuum mechanical treatments of brittle solids based on Griffith's
theory$^{\ref{Griff}}$ are quite widespread and accepeted in 
scientific papers, and
have been relatively successful for practical and engineering
purposes.$^{\ref{Lawn}}$
There are, however, a couple of problems being associated with 
classical fracture mechanics limiting its range of application.
Certain restrictions arise from the assumption that the  
deformations can be conducted in a thermodynamical reversible manner.
Connected to this is the basic assumption that the involved 
fracture processes are thermodynamically reversible too. This is 
usually expressed by minimizing thermodynamic potentials with 
respect to crack/fracture growth, which is a perturbational scheme.
The typical modelling situation is such that some governing field equations
are describing the global state of the deformation (i.e. the free energy
density) according to some boundary/loading conditions while 
the free surface energy {\em field} (intrinsic strength) is classically
assumed to be a constant property throughout the solid material. 
The assumption of a constant intrinsic strength constitutes 
a quite strong idealization for most experimental situations because 
it eliminates the field nature of cohesional properties.
The assumption of a constant theoretical strength is most easily justified 
for {\em homogeneous} materials, i.e., perfect single crystals. 
However, even for such classes of materials there exist certain 
limitations: a) the crystal's cohesion is in general non-isotropic
(variation in crystal strength according crystallographic directions) 
and b) the cohesion is subject of thermal influences (lattice
vibrations) introducing thermal fluctuations into the problem.

The survey becomes more difficult for most solid materials of practical 
interest as they are frequently composite materials. Variations of 
cohesional properties occur already on macroscopic length scales comparable
to those of the microstructural elements. 
There is {\em no} simple 
argument to replace the theoretical strengths of the isolated solid phases 
by some average effective strength in connection with fracture 
problems. This is because the fracture and its associated 
{\em natural geometry} advances on a microstructural level where 
the heterogeneity in grain orientation and strength is present.
We mention this in order to illustrate that the assumption of 
constant cohesion/binding properties throughout a solid 
and therefore the representation employing a homogeneous continuum 
needs careful consideration. 

\begin{figure}[tb]
        \centerline{
        \psfig{file=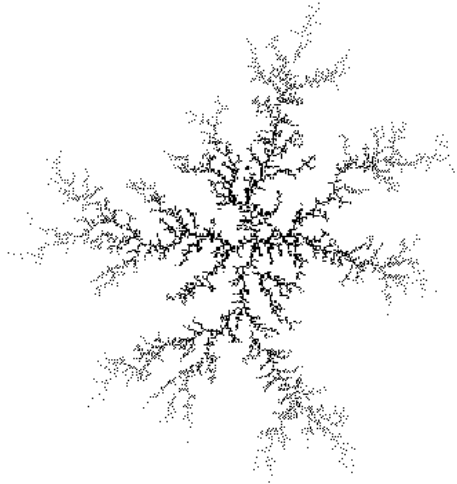,width=0.50\textwidth}
        }
        \caption{Selfsimilar DLA-Cluster in two dimensions consisting 
	of $10^4$ particles. The cluster grows outwards `absorbing' 
	diffusing particles coming from remote distances. 
	The fractal dimension is $d_f = 1.71$. With 
	courtesy of S. Schwarzer (numerical simulation). 
        }
        \label{dla_fig}
\end{figure}

The investigation of geometrical properties 
of irreversibly grown cracks in the presence of disorder has recently 
received considerable attention amoung physicists. Basic ideas 
go back to statistical physics, theory of phase transitions, 
scaling theory, selfsimilarity and self-organized criticality.
These concepts are described in good standard books of condensed matter 
physics, i.e. Chaikin \& Lubensky.$^{\ref{Lubens}}$ \\
Perhaps one of the most well studied phenomena exhibiting 
self-organized criticality is Diffusion Limited Aggregation (DLA), 
theoretically first implemented by Witten \& Sander.$^{\ref{Witten}}$
Its outline is quite simple and it is well suited to exemplify 
geometrical selfsimilarity.\\
Particles being small enough to be subject to Brownian motion diffuse 
in a large gas or fluid reservoir. The particle concentration is 
assumed to be very low (infinite dilution), i.e., there are at each 
time step precisely one diffusing particle and one cluster where 
the diffusing particle is allowed to be attached to. If the diffusing
particle touches the particle-cluster it sticks to it forever and a 
new diffusing particle is started far away from the cluster.
The process is called diffusion limited because the particle diffusion
is the rate limiting step in this reaction scheme.
Figure $\ref{dla_fig}$ shows such a DLA cluster.
The cluster has a dendritic structure exhibiting branches on all 
length scales between particle and cluster length scale. This supposes
that the average number $N$ of particles belonging to the cluster 
follows a power law in cluster radius $R$, i.e.,
\begin{equation}
N\sim R^{d_f}.
\label{equation:df}
\end{equation}
The exponent $d_f$ is called the {\em fractal dimension} of the 
cluster. 
The value for $d_f$ is in general always smaller equal the embedding 
dimension; in the above example $d_f=1.71< d=2$. 
The physical meaning of a power law relation Eq.(\ref{equation:df}) is
that the structure does not contain any natural length scale 
and that it looks similar to itself (selfsimilar) under 
a change of scale. 
The reader interested into concepts and applications of Fractals
is referred to References \ref{Feder} and \ref{Bunde}.
Nowadays it is widely believed that the value of the fractal dimension
does only depend on the symmetries of the governing equations but not 
on particularities such as the particle shape or the employed discretisation 
scheme in numerical simulations.
Deviations from scaling behavior are anticipated for systems of 
finite size.\\

\begin{figure}[tb]
        \centerline{
        \psfig{file=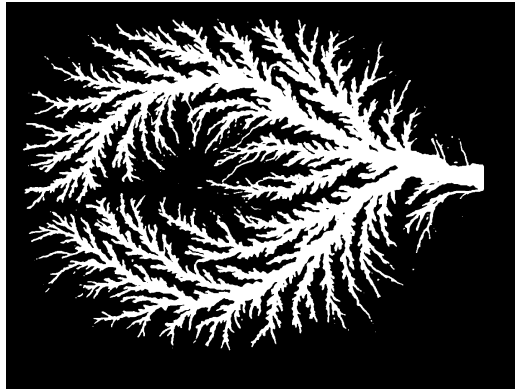,width=0.50\textwidth}
	}
        \caption{Lichtenberg figure in an epoxy plate. The figure 
	shows the electrical discharge pattern towards a notch at the right
	side of the plate.$^{\ref{licht}}$
	}
\label{lichtenberg_fig}
\end{figure}

\begin{figure}[tb]
        \centerline{
        \psfig{file=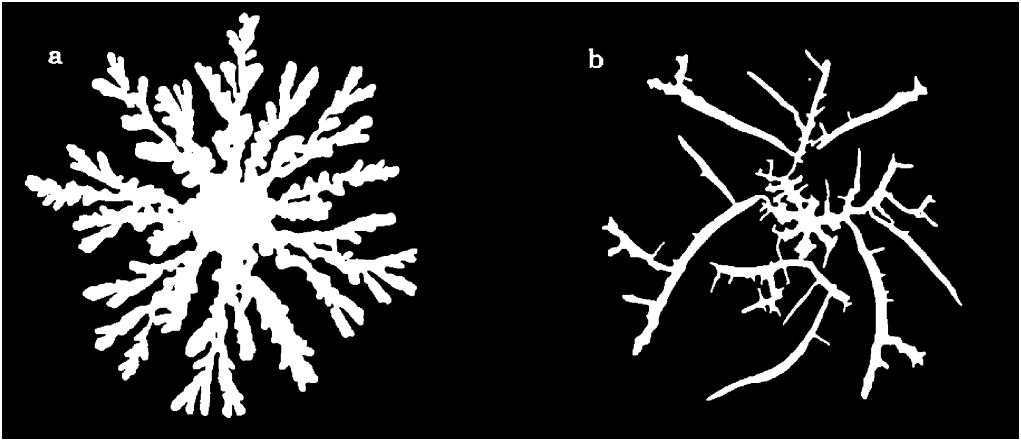,width=1.0\textwidth}
        }
        \caption{Fractal structures obtained after injecting water 
	into Hele-Shaw cells containing clay pastes. 
	(a) Viscous Fingering in a thin liquid paste (solid/liquid =
        0.08). The fractal dimension is $d_f = 1.65$.
	(b) Hydraulic Fracturing in a thick paste 
	(solid/liquid = 0.2). The fractal dimension is $d_f = 1.43$. 
	Note that the fracture structure is more spiky than the 
	viscous finger pattern. Both patterns were formed at the same 
	constant injection pressure.$^{\ref{vandam}}$  
        }
        \label{vandamme_fig}
\end{figure}

The main interest in the DLA problem results from the fact that the 
governing equation is the Laplace equation being one of the most 
widespread and simplest equations in physics. 
A diffusing particle sticks with 
probability proportional to $\partial P/ \partial n$ to the
clustersurface with $P$ being the solution of the diffusion equation.
The sticking-probability is highest near the clusters extremities, so 
that the cluster is growing at a higher rate at the tips than in 
its interior. The tips screen the cluster interior to a certain
amount leading to a diluted dendritic structure.
Other examples belonging to the same {\em universality class} as DLA
are the dielectric breakdown (DBM) (see Fig.~\ref{lichtenberg_fig}), 
the viscous fingering (VF) (see Fig.~\ref{vandamme_fig}a)
and the Stefan problem of 
solidification.$^{\ref{growth}, \ref{Kess88}}$ 
All this problems lead experimentally and numerically to similar 
geometrical patterns.


There has been considerable discussion whether fracture processes 
in presence of fluctuations would lead to DLA-like breakdown pattern 
too. The conducted investigations do, however, not support this idea 
in an unambiguous way. 
Rather the structure of the basic elastic equations (Lam\'e or Cosserat
equations) {\em cannot} be reduced to a Laplace type equation even 
in limiting cases for the elastic constants.
The experiments of Van Damme and collaborators on twodimensionally confined 
clay pastes display a sharp transition between DLA morphologies
(viscous fingering, compare Fig.~\ref{vandamme_fig}a) 
and the much more spiky and diluted 
hydraulic fractures, e.g. see Fig.~\ref{vandamme_fig}b.

Further below we will present numerically obtained results mainly concerned
with the problem of hydraulic fracturing of disordered materials in 
two dimensions. \\
We will first consider the case in which the model material 
contains only elastic interactions. We will investigate the cases 
of a constant injection (borhole-) pressure and of a constant fluid 
influx. \\
In a later chapter we consider a liquid saturated porous 
elastic material being fractured in a similar way. However, in such 
a case is the pressure distribution originating from the pore fluid flow 
directly coupled to the elastic equations.

\section{The Model}
\normalsize
In the following we will outline the employed model. 
We give a brief description of the basic elastic and flow 
equations. 
Thereafter we explain in detail the employed boundary conditions.
After this we demonstrate how heterogeneous 
cohesive properties are taken into account.
Finally we present the breaking rules which contain 
the physics of the
considered breaking process.\\   
\subsection{Pore space}
\label{subsec:pore_space}
There are at least two different ways of implementing  
porosity numerically.
In both cases certain microstructural 
elements (beams, plates or shells) will form a stress carrying 
elastic backbone. One can think, however, of two different realizations 
for the microstructural elements. The elements could be 'made' of a
material which already follows the constitutive laws for porous 
materials, e.g. Biot's 
theory.$^{\ref{Biot41}}$  
This would be most 
satisfying from a continuum mechanical point of view as long as 
no fracturing occurs. However, if {\em fracturing at the pore level} is 
to be taken into account the above mentioned approach does not 
suffice (or even becomes inconsistent) and a micromechanical approach
must be employed. 
The pore space/volume 
has to be modelled explicitely in that case. 
We have focussed on the latter implementation: the microstructural 
elements follow the classical constitutive equations of linear 
elasticity, and the pore space is viewed as the remaining vacant volume. 
The breaking of microstructural elements is associated with 
a coallescence of pore volume under question.
Because the pore volume is modelled explicitely the couplings 
between elastic and flow interactions ought to be derived explicitely 
from the microstructural configuration.\\

In this paper we consider a two-dimensional model-microstructure laying
on a square lattice with properties of the pore space (hydrostatic
pressure field) defined on its 
dual lattice.$^{\ref{dual}}$\\
\subsection{Elastic equations}
\label{subsection:elastic_equations}
\begin{figure}[tb]
	\centerline{
	\psfig{file=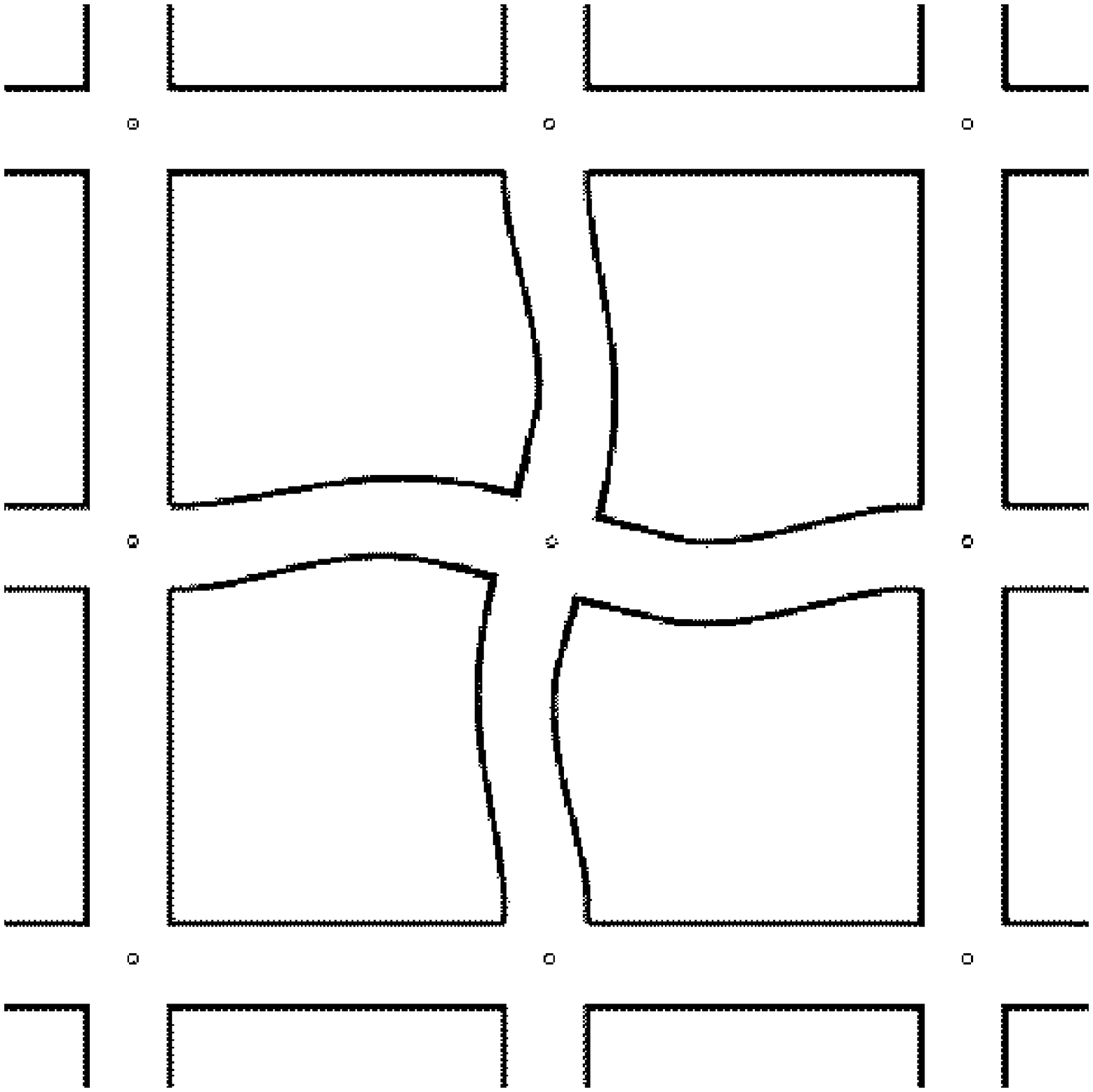,width=0.35\textwidth}
	\hfill
	\psfig{file=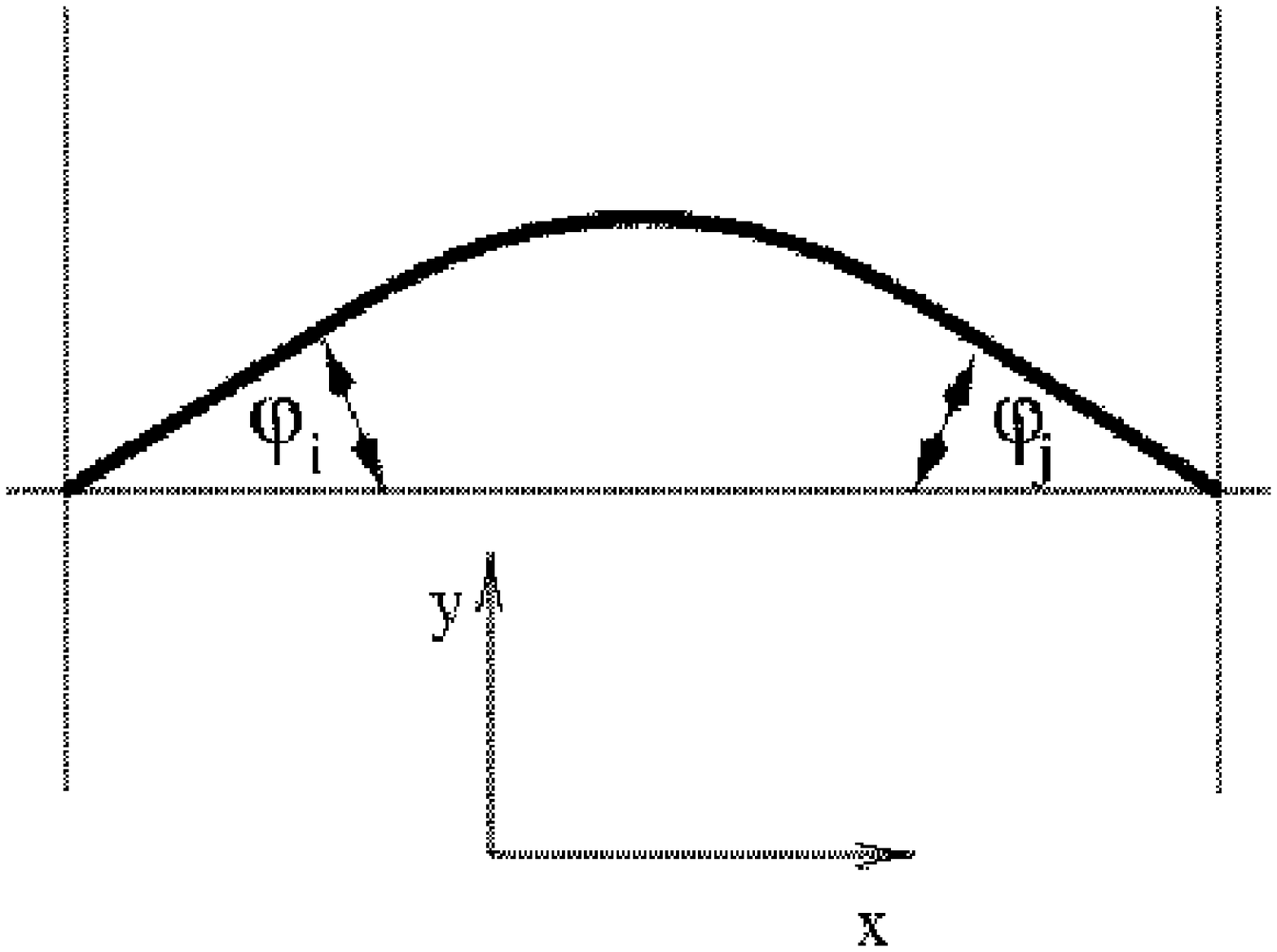,width=0.5\textwidth}
	}
	\caption{Schematic representation of the beam model (left) 
	rotation of one site, (right) beam flexed due to the angles 
	at its extremities.
	}
	\label{model_fig}
\end{figure}

We consider the beam model (as defined in p.~232 of Ref.~\ref{Herr90}) on
a two dimensional square lattice of linear size $L$. This model 
is preferable compared to Hookean spring models
from a microstructural point of view 
because elastic beams due to their finite 
crossection can be physically interpreted as fibers, while Hookean
springs (having zero cross section) can not.  
In the two-dimensional beam model each of the
lattice sites $i$ carries three real variables:
the two translational displacements $x_i$ and $y_i$ and
a rotational angle $\varphi_i $ in the xy-plane. Neighbouring sites
are rigidly connected by elastic beams of length $l$.
The beams are assumed to have the same quadratic cross 
section, $A=d^2$, and
the same elastic behavior, governed by three material and geometry
dependent constants $a=l/(Ed^2)$, $b=l/(Gd^2)$
and $c=12 l^3/(Ed^4)$ with 
$E$ and $G$ being the Young and shear moduli.
With characteristic values for rock materials $E=10^{11}\;Pa$, 
$G= 4\cdot 10^{10}\;Pa$, $l = 10^{-5}\; m$ and $d= 10^{-6}\;m$
the above three constants become 
$a=10^{-4}\;(Pa\;m)^{-1}$, $b=2.5\cdot 10^{-4}\;(Pa\;m)^{-1}$ and 
$c=0.12\;(Pa\;m)^{-1}$.
When a site is rotated ($\varphi_i \ne 0$)
the beams bent accordingly always forming 
tangentially $90^{\circ}$ angles with each other.
In this way local momenta are taken into account.
For a horizontal beam between sites $i$
and $j$ one has the longitudinal force acting at site $j$,
\begin{equation}
\label{equation:longitudinal}
F_j = \alpha (x_i -x_j),
\end{equation}
the shear force
\begin{equation}
\label{equation:shear}
S_j = \beta (y_i - y_j) +{\beta\over 2}l
(\varphi_i +\varphi_j),
\end{equation}
and the flexural torque at site $j$,
\begin{equation}
\label{equation:torque}
M_j ={\beta\over 2}l(y_i -y_j +l\varphi_j) +
\delta l^2(\varphi_i-\varphi_j),
\end{equation}
using $\alpha = 1/a$,
$\beta=1/(b+c/12)$ and $\delta=\beta(b/c+1/3)$.
The corresponding equations for vertical beams are similar. 
The outline of the beam network is displayed in Fig.\ref{model_fig}. 
We will not consider inertial or gravity forces, however, the flow
of a pore fluid amounts in general to non-vanishing body 
forces. 
If there is a pressure drop $p_2 - p_1$
across a beam (pore wall) the resulting line force density
can be replaced by a shear force dipole, $S=(p_1 - p_2) ld /2$, 
acting at both ends of that beam. 
This follows directly from the equilibrium conditions for beams. 
Hence the dipoles couple the pressure field in form of body 
forces to the mechanical equilibrium conditions.\\

In  mechanical equilibrium the sum over all elastic and 
body forces (torques) acting on site $j$ must vanish.
Thus if the pressure field is known one can calculate the 
displacements.\\

To determine the hydrostatic pressure field one needs to consider flow
equations.
\subsection{Flow equations}
\label{subsection:flow_equations}
The beams exhibit a typical length $l$ and thickness $d$ 
and the liquid saturated equilibrium pore volume is simply $(l-d)d^2$.
This states explicitly the considered material to be a {\em binary}
system, i.e. any volume element is either occupied by elastic material
or by pore fluid.
By varying the ratio of $l/d$ different values for the volumetric porosity
can be achieved (ranging from nearly zero to nearly one).
In order to obtain most simplified flow equations, we make the 
following assumtions: 
a) the pore fluid is incompressible (which appears well 
justified for water) and 
b) the fluid velocity relative to the bulk is proportional to the gradient
of the pressure field (Darcy's relation). 
The continuity equations for solid and liquid mass density lead
to the required pressure equation (in the continuum limit); 
\begin{equation}
\label{equation:pressure}
\Div \Bigr(\frac{k}{\mu}\;\Grad\; p\Bigl) = 
\frac{\partial}{\partial t}\;\Div\; \bf{u},
\end{equation}
with $p$ being the pressure, $\bf{u}$ the displacement vector, 
$k$ the mechanical permeability and $\mu$ the fluid viscosity.
This equation is a Poisson equation for the pressure field
for constant $k$ and $\mu$.
The right hand side of Eq.(\ref{equation:pressure}) represents 
in general a 
time-dependent source term for the pressure field. It characterizes  
the local rate of relative volume change due to elastic
deformations, and hence the net fluid flux itself. In general these 
net fluid fluxes will control the pressure field's time-evolution.
However, in this work we will only consider the stationary case. 
The physical motivation for this is the observation, that the
characteristic internal relaxation time $\tau =\mu/E$ 
for water-rock systems is typically 
of order $10^{-6}\;s$. 
It appears justified to assume that the time periods between 
sucessive breaking events are much larger than the internal
relaxation time $\tau$, and
hence that breaking events are triggered by stationary pressure 
and displacement distributions. 
In connection with Eq.(\ref{equation:pressure}) this implies to
balance the net fluid flux for a given cell $i$ to zero (Laplacian);
\begin{equation}
\label{equation:pressure_discrete}
\frac{1}{\mu}
\sum_j k_{ij}( p_i - p_j)\;\;\; 
+\mbox{boundary conditions} = 0.
\end{equation}
Here $k_{ij}$ denotes the local permeabilities across the 
beams (pore walls).\\

The pressure equation Eq.(\ref{equation:pressure_discrete})
is solved without referring to any mechanical
properties, which is directly a consequence of explicitely dropping 
all dynamic terms. Its solution for given boundary conditions 
represents the body force under which the elastic
equations, among their own boundary conditions, are solved. 
Formally the corresponding continuum problem is analogous to 
the standard problem of thermoelasticity$^{\ref{Boley}}$ which was
already noted by Biot$^{\ref{Biot41}}$. 
From the resulting stresses/strains and breaking thresholds (see
Sec.\ref{subsection:heterogeneities}) we then determine which 
element(s) are broken next.\\

Here we would just like to mention what happens to the pressure 
equations, when a beam is broken.
Because the beams represent the pore walls, and
because the fluid is assumed to be incompressible the pressure
within a connected crack has to be constant. In this sense a crack 
is a macro-pore. Numerically this can be accomplished in different  
ways. We have chosen to set the permeabilities for broken beams 
very high, i.e. $k_{ij}=10^{9}$. Hence the pressure drop
along a crack is negligible small.\\

\subsection{Boundary conditions}
\label{subsection:boundary_conditions}

The results presented in this review are mainly for hydraulic
fracturing conditions in impermeable$^{\ref{Tzsch94},\ref{Tzsch95}}$ 
and permeable$^{\ref{Wangen}}$ materials. The 
only exception is Sec.~\ref{section:Rupture_without_Disorder} where 
we compare results for tensile and hydraulic loading
configurations.$^{\ref{Tzsch95-2}}$  
The corresponding tensile boundary conditions 
are therefore discussed in Sec.~\ref{section:Rupture_without_Disorder}.

Let us make few remarks about terminology.
A boundary will be called `free' if its 
normal stress component vanishes and will be called (hydrostatically)
`loaded' if the normal stress component follows some known non-zero  
stress distribution (von Neuman condition).
A boundary is called `fixed' or `clamped' if the displacement 
values are assigned to the boundary (Dirichlet condition). 
A particularity in calculations are the sometimes employed 
`periodic' boundary conditions for displacements/stresses. 
They enforce periodic displacement/stress field solutions.
Periodic boundary conditions are especially useful in numerical calculations 
when the asymptotic behavior for infinite systems is of interest.   
We will encounter them in Sec.~\ref{section:Rupture_without_Disorder}.

For most physical fracture problems the boundary conditions 
for the displacements
are von Neuman conditions, or mixed boundary conditions.
In presence of an additional flow equation boundary conditions for 
the fluid pressure field need to be specified. In our case 
homogeneous Dirichlet conditions are sufficient, i.e., the hydraulic 
crack pressure, $p$, and an equilibrium pressure, $p_0$, for the 
external boundary.

In Sec.~\ref{sec:Hydraulic-Model} and Sec.~\ref{sec:annealed}
we describe simulations where 
the hydraulic crack is `loaded' and propagated by a constant crack
pressure. The crack pressure does neither vary in space nor in time.
The considered external boundaries (lattice edges)  are either free
or periodic.

A, for practical purposes, somewhat more realistic boundary condition
propagating a
hydraulic crack is a constant fluid injection rate, $q_{in}$.
To keep things tractable we only consider an incompressible hydraulic 
fluid. The boundary condition 
for an impermeable crack, Sec.~\ref{subsec:impermeable}, is a
particular case of the more general permeable crack boundary
condition, see below. 
A constant fluid injection rate $q_{in}$ corresponds to a certain 
driving pressure difference, $p-p_0$, where $p$ denotes the hydraulic  
pressure within the crack and $p_0$ corresponds to an equilibrium 
pressure. We have used $p_0 =0$. The hydraulic pressure does depend 
on $q_{in}$ and the crack opening volume. For permeable 
cracks/materials $p$ also depends on the hydraulic losses.
In the following we will outline how to determine the hydraulic crack
pressure $p$ for given injection rate $q_{in}$ and crack structure 
(contour) $\partial\,V$.
 
Because for an incompressible fluid the conservation of fluid mass 
directly implies a conservation of fluid volume, it is possible 
to formulate a continuity equation for the crack opening volume $V$,
\begin{equation}
\frac{\partial\, V}{\partial\,t} +
\int_{\partial\,V} {\bf v} \cdot d\,{\bf A} = q_{in}.
\label{equation:nice_equation}
\end{equation}
Here $\bf{v}$ denotes the Darcy field integrated over the crack
surface $\partial\,V$ with surface element $d\,\bf{A}$. As source 
term appears the rate of injected (fluid) volume, $q_{in}$. Because 
the Darcy vectors are linear funtionals of the pressure, 
${\bf v}= -k/\mu \cdot \Grad p$, and the crack opening volume is a linear 
fuctional of displacements, $V=\int_{\partial\,V}{\bf u}\cdot d\,{\bf
A}$, and because all considered equations are linear, a scaling of the 
form $p=\lambda p^*$ implies ${\bf v}=\lambda {\bf v^*}$, 
${\bf u}=\lambda {\bf u^*}$ and $V=\lambda V^*$.
Note that such a scaling is in general {\em impossible} if one 
considers Eq.~(\ref{equation:pressure}) instead of the 
Laplacian, $\Delta\,p = 0$, as flow equation.

There exists an interesting application of
Eq.(\ref{equation:nice_equation}) for undercritical, non-growing cracks. 
With $V^*$ and ${\bf v^*}$ depending {\em only} on the constant crack 
contour $\partial\,V$ and crack pressure $p^*$
one obtains a first order differential 
equation for the scaling factor $\lambda (t)$,
\begin{equation}
\frac{d\,\lambda}{d\,t} \cdot 
\int_{\partial\,V}{\bf u^*}\cdot d\,{\bf A} +
\lambda \cdot \int_{\partial\,V}{\bf v^*}\cdot d\,{\bf A} = q_{in},
\label{equation:lambda_equation_continuum}
\end{equation} 
with general solution,
\begin{equation}
\lambda (t)= \lambda_{\infty}+ (\lambda_0 - \lambda_{\infty})\cdot
e^{-t/\tau},
\label{equation:lambda_solution}
\end{equation}
using

\begin{equation}
\lambda_{\infty} = \frac{q_{in}}
{\int_{\partial\,V}{\bf v^*}\cdot d\,{\bf A}},\;\;\;
\lambda_0 = \frac{p_0}{p^*}, \;\;\;
\tau = \frac{\int_{\partial\,V}{\bf u^*}\cdot
d\,{\bf A}}{\int_{\partial\,V}{\bf v^*}\cdot d\,{\bf A}}.
\label{equation:lambda_constants}
\end{equation}

The important point is that one can choose an {\em arbitrary} boundary
value $p^*$ within a calculation; 
the resulting $p=\lambda (t)\cdot p^*$ will be unique. 
However, one needs to calculate the contour integrals
in Eq.~(\ref{equation:lambda_constants}) in order to do the rescaling.

The magnitude $\lambda_{\infty}$ does depend only on the
injected fluid flux $q_{in}$ and the total fluid flux through the 
crack contour. It does {\em not} depend on elastic
properties.
The typical time scale $\tau$, above which the crack pressure and the 
crack opening volume tend to approach their asymptotic values, is 
given by the ratio of crack volume
and fluid volume flux through the crack surface. It does 
{\em not} depend on the injected fluid flux.
Again we would like to stress that the crack contour is assumed to 
be constant. However, Eq.(\ref{equation:lambda_solution}) is quite
general in the sense that all dependencies on the actual crack
contour have been implicitly taken into account within the constants 
$\lambda_{\infty}$ and $\tau$. 
Equation (\ref{equation:lambda_solution}) should therefore hold 
for arbitrarily shaped cracks.

There are two limiting cases for the practical important situation 
$\lambda_0 \approx 0$. For large times, $t\gg \tau$, the scaling 
factor $\lambda (t) \to \lambda_\infty$ becomes constant and so do 
the crack pressure $p_\infty= q_{in}\tau \cdot (p^*/V^*)$ and the crack 
volume $V_\infty= q_{in}\tau $.
The other limiting case is for small times, $t\ll \tau$,  where the 
scaling factor becomes linear in time, 
$\lambda (t)= \frac{q_{in}}{V^*}\cdot t$ with corresponding 
pressure $p= q_{in}\frac{p^*}{V^*}\cdot t$ and volume
$V= q_{in}\cdot t$.
The latter case also holds for impermeable materials because 
$\lim_{k\to 0}\lambda (t)= q_{in}/V^* \cdot t$.

Equations (\ref{equation:nice_equation}) and 
(\ref{equation:lambda_solution})
hold for {\em arbitrary}
crack contours, i.e., branched or ramified cracks.

Equation \ref{equation:nice_equation} has a
corresponding discrete update scheme. One just needs to be 
careful with the time derivative in
Eq.~(\ref{equation:nice_equation}) because the scaling 
factors $\lambda$ are in general different before and after the 
time increment, $\Delta\,t$,
\begin{equation}
\lambda (t+\Delta\,t) =
\lambda(t)\cdot \Bigl( 1 - \frac{D^*}{V^*}\Delta\,t \Bigr) +
\frac{q_{in}}{V^*}\Delta\,t .
\label{equation:lambda_equation_discrete}
\end{equation}
Here $D^* = \int_{\partial\,V}{\bf v^*}\cdot d\,{\bf A}$ and 
$V^* = \int_{\partial\,V}{\bf u^*}\cdot d\,{\bf A}$ denote the 
crack contour integrals/sums for the Darcy flows and displacements
respectively, as obtained from the solution of the 
boundary value problem at time $t$ with boundary value $p^*\ne 0$.
As soon as the crack growths and its boundary changes one needs to
recaculate the flow and the elastic equations in order to obtain the
corresponding $D^*$ and $V^*$. The ratio $D^* / V^*$ being the inverse
of a typical time will therefore change during the crack growth process.

We summarize:
\begin{enumerate}
\item\label{a} 
Choose an initial crack boundary.

\item\label{b} 
Set the initial value $\lambda (0)$, i.e., $\lambda (0)=0$ (closed crack).

\item\label{c} 
Solve the pressure equation
Eq.~(\ref{equation:pressure_discrete}) for an arbitrary hydraulic
pressure $p^*$ (boundary value for the current crack boundary).

\item\label{d} 
Insert the corresponding pressure gradients into the 
elastic equations 
Eqs.~(\ref{equation:longitudinal})-(\ref{equation:torque}) as
inhomogeneities and solve the elastic equations.

\item\label{e} 
Calculate from the obtained elastic solution the stress/force 
distribution $\sigma^*$. 

\item\label{f} 
Calculate from the elastic solution the crack opening volume
$V^*$ and from the pressure solution the total Darcy flux $D^*$.

\item\label{g} 
Update $\lambda$ according to 
Eq.(\ref{equation:lambda_equation_discrete}).

\item\label{h} 
Use the updated stresses/forces, $\sigma = \lambda \sigma^*$, 
in the breaking rule (see Sec.~\ref{subsection:breaking_rules}) 
and break the corresponding element(s).

\item\label{i} 
Repeat steps \ref{c}-\ref{h} until a stopping criterion is fulfilled.

\end{enumerate}

\subsection{Heterogeneities}
\label{subsection:heterogeneities}
The problem of fracture processes in elastic media has undergone much
attention in the last years, in particular
the role of disorder in such processes.$^{\ref{Sahim86}-\ref{p10}}$
An important question of interest is how a
variation in the {\it local} properties affects the {\it global}
elastic properties in comparison to ideal homogeneous media.
It was recognized that elastic properties could
considerably be changed and therefore interest was attracted
how failure mechanisms like fracture, rupture or damage processes
are influenced, changed or possibly controlled by 
disorder.

In the literature are two essentially different ways to take the
disorder into account.$^{\ref{Herr90}}$ 

The first possibility is to use exclusively the strain/stress field 
in order to calculate the probability for
breaking a local constitutive element. 
This is known as stochastic disorder. Prominent {\em scalar} examples
for this kind of disorder are known in the literature as
Diffusion Limited Aggregation (DLA), Dielectric Breakdown or Viscous
Fingering.$^{\ref{growth}, \ref{Kess88}}$ 
We will present results for stochastic disorder in 
Sec.~\ref{sec:annealed}.

A second different kind of disorder is to initially assign a
statically set of random values to certain 
material specific constants (structural disorder). Once these 
constants like strength, bending thresholds, elastic properties and
others have been chosen
the system behaves in a fully deterministic way. 
Corresponding results are presented in Sec.~\ref{sec:quenched}.

\subsection{Breaking rules}
\label{subsection:breaking_rules}
Lattice fracture models like the beam model
are spatially discrete. This is because one would like to have
direct access to
each constitutive elastic element. The
medium is represented by a lattice and nearest neighbouring lattice
points are connected by constitutive elastic elements called
bonds. The bonds represent interactions not on an atomistic scale
but rather on a
mesoscopic scale, for example the elastic coupling between two
adjacent grains where the grains themselves are linear
elastic bodies. 
New features are the definitions of physically
motivated ''selection" and ''breaking" rules for lattice bonds.
Either a bond is broken according to some
breaking rule and stays broken for the future or it remains intact
and is further susceptible for the breaking process.
The simulation ends when
the elastic lattice fully breaks apart.
Moreover, one has to define in the selection rule those bonds which
could break
in principle at a given stage of rupture. In single
crack models only the surface bonds of the crack are allowed to
break.
However, in the many crack version in general all bonds are
eligible. This allows in principle the nucleation of several microcracks
in the system. We have used for our simulations the first kind of 
selection rule. 
In a recent
published work$^{\ref{p8}}$ it was shown
that if the disorder for the breaking thresholds is not to weak
there are two different regimes in the macroscopic
breaking characteristic.
For small macroscopic strains first the weakest bonds are broken
without leading to catastrophic failure. The breaking process
is stable in this region and one has to increase the external applied
stress in order to proceed breaking bonds. In finite systems
however, this disorder controlled regime must end at the
breaking point. The breaking process localizes, 
becomes unstable and a single
large crack controlls the entire breaking. It was shown that for
different probability distributions of breaking thresholds and
different loading conditions two
nontrivial possibly universal 
exponents are sufficient to rescale the macroscopic breakdown
characteristics with respect to the system size. 
Breaking rules which determine the breaking conditions
are crucial in lattice models
because one has to put additional information about the fracture
{\it mechanism} into the model which could not be obtained from
elasticity theory. 
The employed breaking rules are described in the corresponding 
sections Sec.~\ref{section:Rupture_without_Disorder}, 
Sec.~\ref{sec:annealed}, and Sec.~\ref{sec:quenched}.

\section{Fracturing without disorder}
\label{section:Rupture_without_Disorder}
\normalsize
We investigate both the case of uniaxial 
loading due to remote tensile forces and  
the loading due to a  
hydrostatic pressure acting from the interior on the crack surface.  
The former situation is typically encountered when a material is 
pulled at its outer faces, while the latter one can be 
found in engineering applications of hydraulic 
fracturing.$^{\ref{s3-2a}}$
We will neglect inertial forces, assuming a 
sufficiently slow fracture process. 
The cohesive properties of the beams are considered to be 
homogeneous, i.e.\  
the beams behave linear-elastic up to their theoretical strength
$F_{th}$ (cohesion force), above which they irreversibly 
break (zero elastic constants).
At the beginning of each simulation we break one vertical beam 
located at the center of the lattice. This is the initial crack.
Corresponding to the employed boundary conditions described below  
we calculate the internal force distribution $F_{ij}$ of beams 
connecting sites $i$ and $j$ using a conjugate gradient 
method.$^{\ref{s3-6}}$ 
From this longitudinal forces we
determine and break the most over stressed beam, i.e. 
the beam carrying the highest force, 
$F_{i_0j_0}= \max_{\{ij\}}F_{ij}$. The maximum force 
allows us to calculate a scaling factor $\lambda = F_{th} /
F_{i_0j_0}=\sigma_{th}/\sigma_{i_0j_0}$ which in  
turn is used to determine the
macroscopic breaking stress $\sigma_c$
or breaking pressure $P_c$ (measured in units of $\sigma_{th}$)
necessary to fulfill the 
local fracture criterion 
$\sigma_{i_0j_0}=\sigma_{th}$.
Such scaling is possible due to the linearity of the employed 
elastic equations.
Breaking the beam connecting sites $i_0$ and $j_0$ destroys 
the balance of forces at those sites and one has to relax
the system to its new equilibrium configuration. Then the 
above steps are repeated until the lattice breaks apart. 

It is well known that the highest 
stress enhancement factors occur at the crack tips 
for the pressure as well as 
for the tensile problem. Because we consider only  
homogeneous cohesion forces the terms `most over stressed' and 
`highest stress' are equivalent and the cracks grow linear 
in direction of the highest local tensile force. In our case 
the direction of crack growth is 
the x-direction because of the uniaxiality
of imposed boundary conditions, see below.

\subsection{Tensile fracturing}
\label{sec:Tensile-Model}
To impose an external strain we attach at the bottom of the 
lattice a 
zeroth line on which for all sites $x_i=y_i=\varphi_i=0$ 
are fixed, 
and on the top we attach a $(L+1)$st line on which all sites 
have the same fixed values $x_i=\varphi_i=0$ and $y_i=1$. 
With this boundary conditions the external displacement in 
y-direction
is fixed to unity (Dirichlet boundary conditions) and one can 
imagine them as being represented by rigid bars attached 
at the top and the bottom of the lattice.
For the boundary conditions in x-direction we consider 
periodic as well as free lattice boundaries.
The externally applied stress, $\sigma_{0}$, necessary to maintain 
a unity 
displacement between the two rigid bars can be easily calculated 
by summing up all $y_i$ displacements over the first line, 
$\sigma_0 = {\alpha\over{A\,L}}\sum y_i$, where $\alpha$ is the 
longitudinal force constant, $A$ the cross section of the beam and 
$L$ the (dimensionless) lattice size. Employing the above defined 
scaling factor $\lambda$ we express the externally applied 
breaking stress, $\sigma_c$, in the form,
\begin{equation}\label{eq:Tensile}
\sigma_c = \lambda \sigma_0, 
\end{equation}
or in dimensionless form, 
\begin{equation}\label{eq:Tensile-Stress}
{\sigma_c\over\sigma_{th}}= {1\over F_{i_0j_0}}
{\alpha\over L}\sum y_i.
\end{equation}
With this notation the external breaking stress $\sigma_c$ 
is just the
stress necessary to yield the local breaking condition 
$\sigma_{i_0j_0}=\sigma_{th}$.

We will use Eq.(\ref{eq:Tensile-Stress}) for our 
finite-size scaling purposes.
It expresses the breaking stress, 
measured in units of cohesion 
stress, in terms of the beam force $F_{i_0j_0}$ at the `crack 
tip' and 
the average force ${\alpha\over L}\sum y_i $ on a beam due to 
the global unit displacement.
In general both terms depend on the number of broken beams $N$ 
(crack size) and on the lattice size $L$, 
see Sec.~\ref{sec:Results-Tensile}. 

\subsection{Hydraulic fracturing}
\label{sec:Hydraulic-Model}
Experimentally hydraulic fracturing is encountered
when an incompressible fluid 
is injected under high pressure into an existing crack. 
The characteristics of this particular loading mode is that
the loading of the crack only happens on the crack surface 
itself, but not 
on remote surfaces like in the case of tensile fracturing.  
Recently hydraulic fracturing has been numerically addressed 
by introducing for each 
broken beam a force dipole of strength $F_0$ simulating 
an inner pressure $P_0= F_0/ A$.$^{\ref{Tzsch94}, \ref{Tzsch95}}$
In the present section we will follow this implementation. 
The basic assumptions are a) the pore fluid is highly compressible
(gas), b) the permeability is negligible and c) the injected fluid 
is incompressible.  
Starting from 
one vertical broken beam (zero elastic constants, one 
vertical force dipole)
we calculate the internal stress distribution $F_{ij}$ using 
the conjugate
gradient. Calculating the scaling factor $\lambda$ as 
described above
we determine and break the beam carrying the highest stress, 
$F_{i_0j_0}$. 
Thereafter the lattice has to be relaxed again 
(now two broken beams, two force dipoles) and one repeats the above 
mentioned steps until the lattice breaks apart.
The crack growth is single connected. As always the two vertical beams 
in front of both crack tips are most overstressed and broken.  Hence
the existence of a connected path for the liquid is garantued.  

We have performed calculations for free external boundaries 
in x- and y-direction
as well as for periodic boundaries in 
x-direction and free boundaries in y-direction.
As we only impose stresses the elastic
solution (displacement field) is unique up to a general translation 
and rotation. Without restriction we therefore fix 
the displacement of an arbitrary lattice site to zero.
Analogous to Eqs.~(\ref{eq:Tensile}) and (\ref{eq:Tensile-Stress}) 
one has for the breaking pressure,

\begin{equation}\label{eq:Pressure}
P_c = \lambda P_0,
\end{equation}
and 
\begin{equation}\label{eq:Pressure-Stress}
{P_c\over \sigma_{th}}={F_0\over F_{i_0j_0}},
\end{equation}
respectively.
It should be noted that $\sigma_{th}$ is 
the tensile strength 
and not the compressional strength. 
In fact the stresses at the tips are tensile stresses as the crack 
is hydraulically opened. 

We will use Eq.~(\ref{eq:Pressure-Stress}) 
for the finite-size scaling of the hydraulic problem.

\subsection{Results}
\label{sec:Results}
As explained in Sec.~\ref{sec:Tensile-Model} and \ref{sec:Hydraulic-Model} 
we obtain from our simulations
the macroscopic breaking stresses and pressures, 
$\sigma_c(N,L)$ and $P_c(N,L)$, measured in units of the theoretical 
strength $\sigma_{th}$, 
in terms of the number of broken 
beams $N$ and the employed lattice sizes $L$.
As we are interested in the finite-size scaling properties of the 
breaking stresses ($\Sigma_c =\sigma_c$) and pressures ($\Sigma_c =P_c$)
we introduce two new functions, $\Psi (N)$ and 
$\Phi (N/L)$,
\begin{equation}\label{eq:Scaling-General}
\Sigma_c(N,L)=\sigma_{th}\,\Psi (N)\,\Phi (N/L).
\end{equation}

The function $\Phi (N/L)$ describes the finite-size 
corrections to the scaling $\Psi (N)$ between the breaking 
stress/pressure
and the crack length for given boundary conditions in the infinite 
system. The quantity $N/L$ is for our purposes the most suitable 
scaling variable as it represents the length of a linear crack
measured in units of the employed lattice size.
In general both functions, $\Psi$ and $\Phi$ will depend on the 
employed boundary conditions. 
The dependence of the asymptotic 
scaling $\Psi$ on employed boundary conditions appears to be 
contrary to what is expected from the 
theory of critical phenomena. In fact the asymptotic scalings $\Psi$ for 
breaking stresses and pressures of continua without internal length
scale are identical. However, as we will see below this 
is no longer necessarily true for systems with internal length 
scale. One obtains rather finite-size corrections originating from 
this scale which, depending on the employed boundary conditions, 
can result in very broad, non universal transient regions. 

In order to show numerically that the finite-size scaling 
function $\Phi (N/L)$ exists, one needs to know (or to 
conjecture) the 
explicit form of $\Psi (N)$, i.e. the asymptotic
behavior of $\Sigma_c$ as $L\to\infty$.

We will access the latter information from simple continuum 
mechanical considerations of symmetric elasticity.

\subsubsection{Tensile fracturing}
\label{sec:Results-Tensile}
\begin{figure}[tb]
\centerline{
        \psfig{file=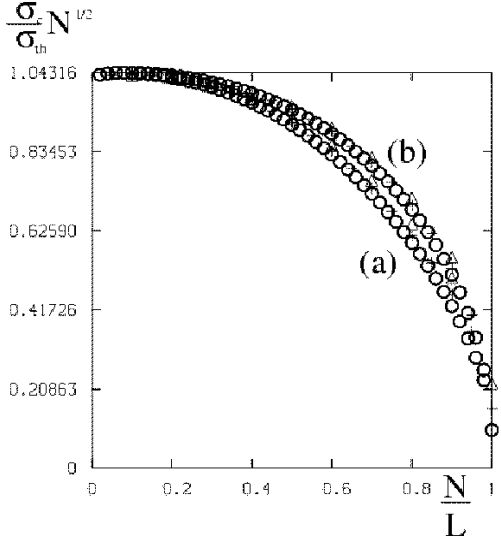,width=0.50\textwidth}
        }
\caption{
         Finite-size scaling for the rescaled breaking stress 
         $\sigma_c/\sigma_{th}N^{1/2}$ as a function of the 
         rescaled crack length $N/L$, (a) for free boundaries in 
         x-direction and (b) for periodic boundaries in x-direction.
         Lattice sizes: ($\triangle$) for $L=20$, ($+$) for $L=40$ 
         and ($\circ$) for $L=100$.$^{\ref{Tzsch95-2}}$
        }
\label{fig:Tension-Scaling}
\end{figure}

Previous continuum mechanical scaling considerations have  
successfully been applied to fracture models.  
It has been shown$^{\ref{Tzsch95-2}}$ that the breaking stress $\sigma_c$
for tensile fracturing of a finite, periodic beam lattice follows 
very well
the known `Tangent-Formula' of fracture mechanics 
$^{\ref{s3-8}, \ref{s3-9}}$,
\begin{equation}\label{eq:Tangent-Formula}
{\sigma_c\over\sigma_{th}}\sim 
N^{-1/2}\sqrt{ {N\over L}\cot{{\pi\over 2}{N\over L}}}.
\end{equation}
By comparison with Eq.~(\ref{eq:Scaling-General}) we see that 
the asymptotic scaling for the breaking stress $\sigma_c$ is given
by $\Psi\sim N^{-1/2}$, a result that originally was 
found by Griffith. 
In Fig.~\ref{fig:Tension-Scaling} we show the finite-size scalings 
for the numerically obtained tensile breaking stresses 
(a) for free boundary conditions and    
(b) for periodic boundary conditions. 
For both curves the data collapse is quite acceptable. While for 
the periodic problem the finite-size correction $\Phi (N/L)$  
follows Eq.~(\ref{eq:Tangent-Formula})$^{\ref{Tzsch95-2}}$, 
no analytical expression for $\Phi$ is known in the case of 
free boundary conditions. 
However, the differences for the breaking stresses at given values of
$N/L$ are less than $0.1\sigma_{th}$, see
Fig.~\ref{fig:Tension-Scaling}, and Eq.~(\ref{eq:Tangent-Formula})
represents a reasonably good approximation for finite solids for many 
practical purposes.

\subsubsection{Hydraulic fracturing}
\label{sec:Results-Pressure}
\begin{figure}[tb]
\centerline{
	\psfig{file=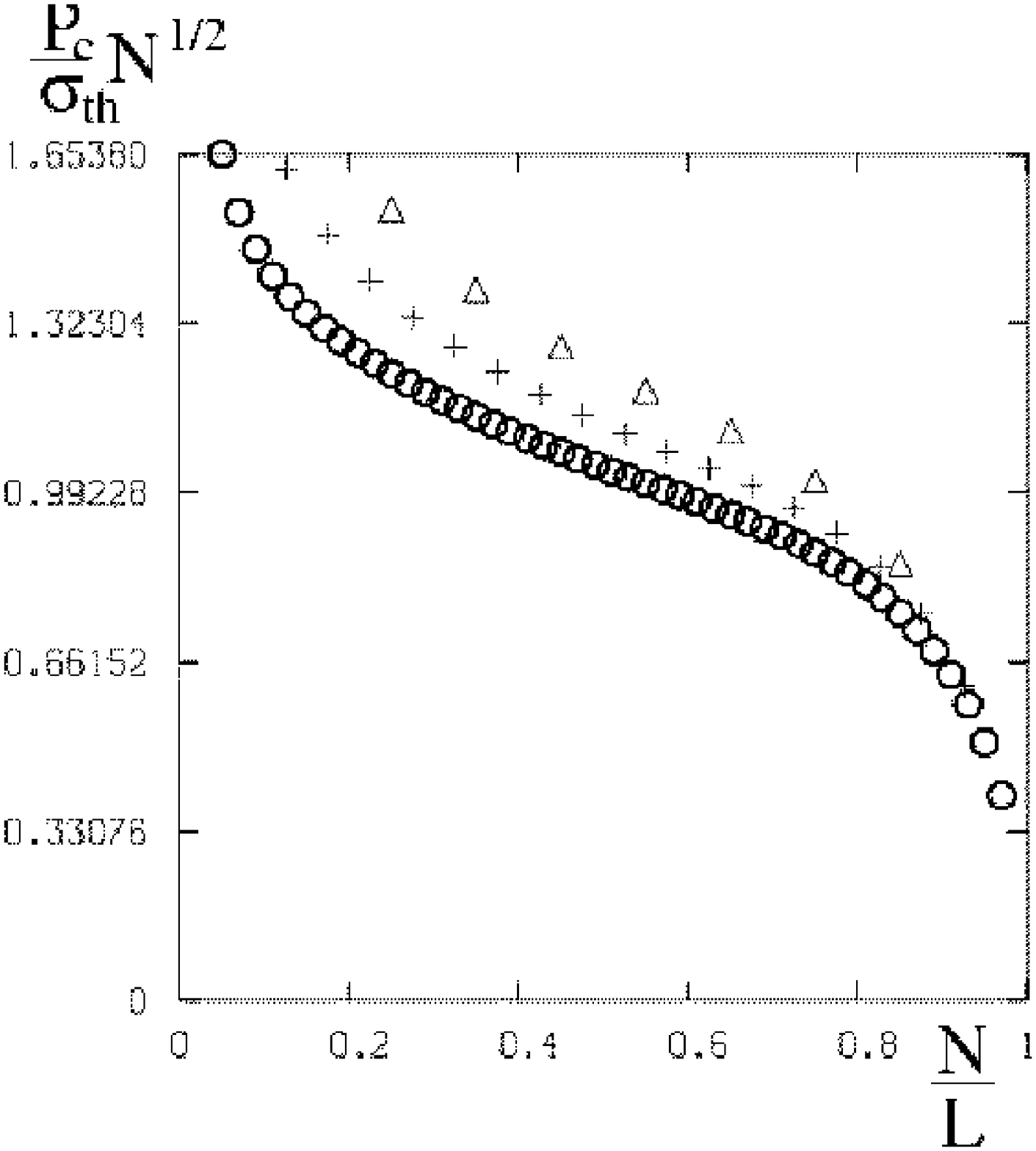,width=0.50\textwidth}
        }
\caption{
         Plot of the rescaled breaking pressure  
         $P_c/\sigma_{th}N^{1/2}$ as a function of the 
         rescaled crack length $N/L$ for periodic boundaries 
         in x-direction. In contrast to
         Fig.~\ref{fig:Tension-Scaling} we do not find any
         data collapse, indicating that $\Psi (N)\sim N^{-1/2}$ 
         in Eq.~(\ref{eq:Scaling-General}) is not the  
         appropriate asymptotic expression for the breaking pressure.         
         Lattice sizes: ($\triangle$) for $L=20$, ($+$) for $L=40$ 
         and ($\circ$) for $L=100$.$^{\ref{Tzsch95-2}}$
        }
\label{fig:Pressure-wrong-Scaling}
\end{figure}
In general analytical solutions of crack problems are 
difficult to obtain and even for two-dimensional 
problems in infinite, semi-infinite or  
periodic domains one is often faced with complicated integral
equations. The reader interested in this is referred to  
Muskhelishvili$^{\ref{s3-10}}$ and Sneddon$^{\ref{s3-9}}$.

It has been shown analytically that the breaking pressure $P_c$ 
of a semi-infinite two-dimensional continuum with periodic boundary 
conditions is given by Eq.~(\ref{eq:Tangent-Formula})  
where one has to replace $\sigma_c$ by $P_c$.$^{\ref{s3-9}}$ 
We show in Fig.~\ref{fig:Pressure-wrong-Scaling} the analogous plot 
to Fig.~\ref{fig:Tension-Scaling} for the breaking 
pressure. One does not obtain a data collapse, indicating that 
$\Psi\sim N^{-1/2}$, see Eqs.~(\ref{eq:Scaling-General}) and
(\ref{eq:Tangent-Formula}), is 
{\em not} the scaling of the breaking pressure in the asymptotic limit 
of infinite lattices.

This result is somewhat unexpected, because the continuum limit
predicts identical scaling relations for breaking stresses and
pressures and on the other hand we find 
good agreement between lattice and continuum scalings 
for the tensile problem.

However, the proposed model for hydraulic fracturing is defined on 
a lattice and not in a continuum. As lattices always show additional
structure, higher order gradient terms of the 
continuum displacement field
appear in the elastic solution, which have no counterparts in 
simple continua.$^{\ref{s3-11}, \ref{s3-4b}, \ref{s3-4d}}$
 
In the following we present an argument why the lattice 
finite-size scalings 
for the pressure and tensile problem are different.

In the tensile problem the loading forces are acting at the 
lattice boundaries {\em remote} to the stress free crack surfaces. 
Therefore local stresses close to the crack tips are carried  
through nearly the whole elastic volume. 
Contrary to this in the hydraulic problem loading 
only happens due to force 
dipoles acting at the crack surfaces {\em close} to the crack tips. 
One therefore might expect that the breaking pressures $P_c$ show a more 
pronounced `(micro structure) lattice behavior' than the breaking 
stresses $\sigma_c$. For very large cracks these differences should 
diminish as the discrete loading approaches a force density. 

In order to extent a linear crack of length $N$ in an infinite continuum
one has to impose a breaking pressure $P_c \sim N^{-1/2}$.
However, employing a {\em single double force} of magnitude $F_c$
one finds, $F_c\sim N^{1/2}$ for the critical loading force 
(related Boussinesq problem).$^{\ref{s3-12}}$ 

We argue now that for a single crack in an infinite disk loaded by 
equidistant force dipoles the asymptotic scaling for $\Psi$
is equivalent to that of an infinite lattice.
Knowing $\Psi$ we plot the finite-size scalings  
for the investigated lattices.

The asymptotic continuum scaling for the breaking pressure (as
calculated from equidistant force dipoles)
is most easily obtained using the complex stress 
function of Westergaard.$^{\ref{s3-12},\ref{s3-8}}$
We find for the asymptotic scaling in an infinite disk,
\begin{equation}\label{eq:improved-scaling}
\Psi (N)\approx {\pi\over 2}{N^{1/2}\over 1+N\,f(N)},
\end{equation}
with
\begin{equation}
f(N)=\sum_{k=1}^{(N-1)/2}{1\over\sqrt{({N+1\over 2})^2 -k^2}},
\quad\mbox{$N$ odd}.
\end{equation}
The sum $f(N)$ converges rapidly towards $\arcsin({N-1\over N+1})$ 
and for large $N$ towards $\pi /2$. 
Hence in the limit of a large number of force dipoles we obtain 
the aforementioned asymptotic Griffith scaling for the breaking pressure, 
$P_c / \sigma_{th}=  N^{-1/2}$.

\begin{figure}[tb]
\centerline{
	\psfig{file=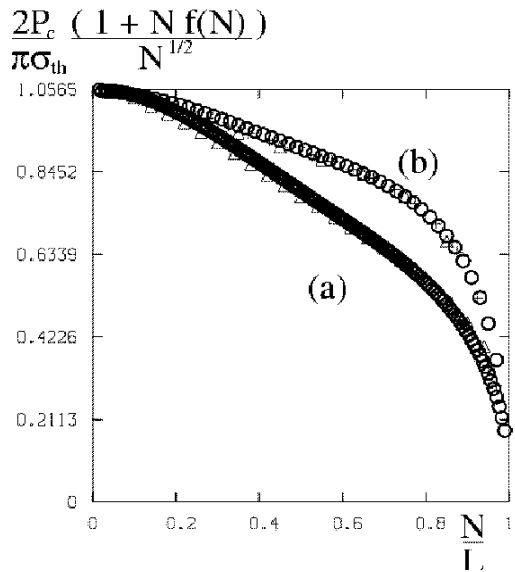,width=0.50\textwidth}
        }
\caption{
         Finite-size scaling for the rescaled breaking pressure  
         $P_c/\sigma_{th}\Psi^{-1}(N)$ with 
         $\Psi^{-1}(N)={2\over\pi}{1+N\,f(N)\over N^{1/2}}$, see 
         Eq.(\ref{eq:improved-scaling}), as a function of the 
         rescaled crack length $N/L$; 
         (a) for free boundaries $(\triangle):\, L=50$, $(+):\, L=100$,
         $(\circ):\, L=300$; (b) for periodic boundaries 
         $(\triangle):\, L=20$, $(+):\, L=40$, $(\circ):\, L=100$.
         Note the data collapses.$^{\ref{Tzsch95-2}}$
        }
\label{fig:Pressure-Scaling}
\end{figure}

In Fig.~\ref{fig:Pressure-Scaling} we show the resulting 
finite-size scalings for the breaking pressure, based on 
Eqs.~(\ref{eq:Scaling-General}) and (\ref{eq:improved-scaling}).
In this plot a reasonable scaling is obtained, which proves our 
earlier statement that the asymptotic limits ($L\to\infty$) of the 
breaking pressures for continua and lattices 
are different.

Comparing Fig.~\ref{fig:Pressure-Scaling} with 
Fig.~\ref{fig:Tension-Scaling} one also notes a significant 
deviation from the scaling form Eq.~(\ref{eq:Tangent-Formula}), 
demonstrating a modified correction to scaling behavior. 

\section{Hydraulic Fracturing with stochastic disorder}
\label{sec:annealed}

\begin{figure}[tb]
\centerline{
	\psfig{file=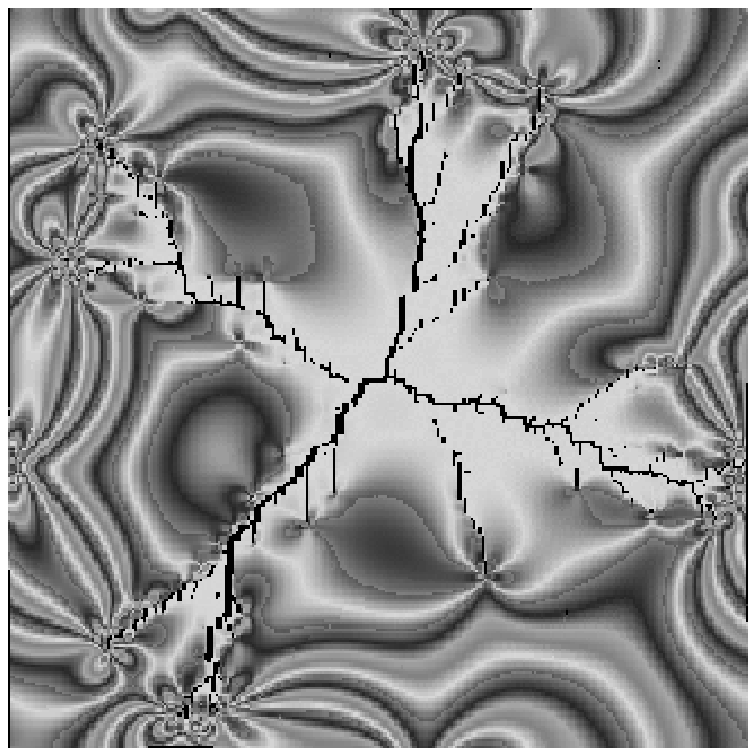,width=0.50\textwidth}
	\hfill
	\psfig{file=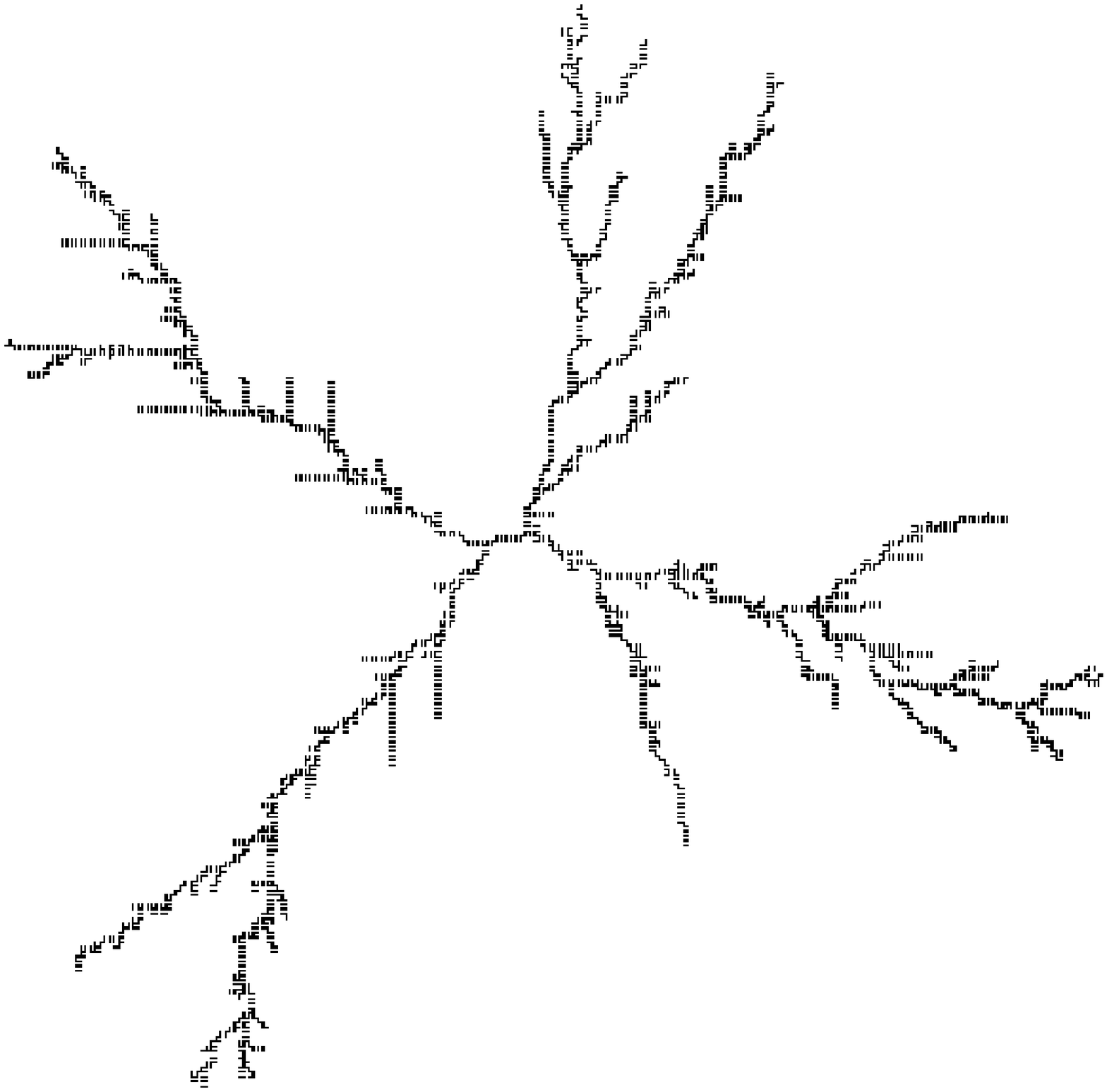,width=0.50\textwidth}
        }
\caption{
	(Left figure): 
	A typical crack obtained with a beam model for hydraulic
	fracturing at constant pressure, on a square lattice of
	$250 \times 250$ with periodic boundary conditions. All beams 
	under tension break with the same probability. The crack 
	consists of $2200$ broken beams. The greyscale represents 
	the difference of the diagonal elements of the hydrostatic 
	stress field. 
	(Right figure): Only the crack structure.$^{\ref{photo}}$
        }
\label{fig:Photoelast}
\end{figure}
\begin{figure}[tb]
\centerline{
	\psfig{file=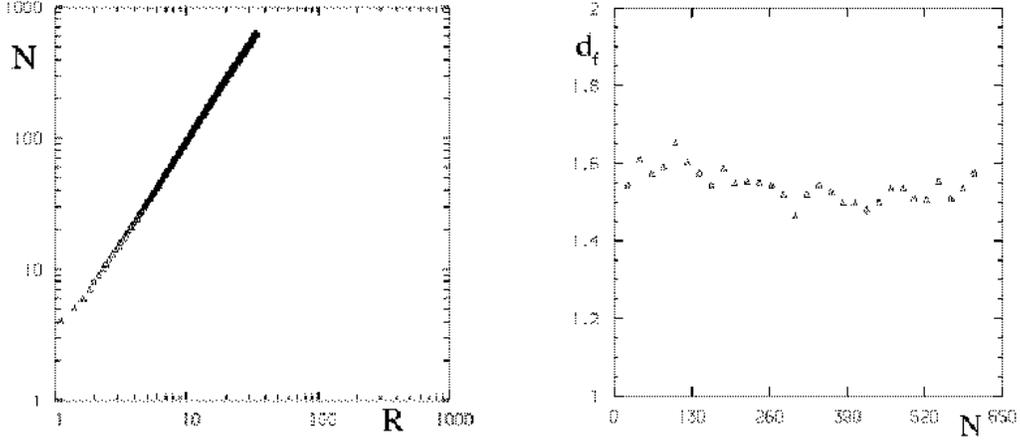,width=1.00\textwidth}
        }
\caption{
	(Left figure): Double-logarithmic plot of the number of 
	broken beams $N$ as a function of the average crack radius $R$ 
	for free boundary conditions. Note the power-law behavior, 
	$N\sim R^{d_f}$, indicating fractal crack growth.
	(Right figure): Plot of the 
	successive slopes $d_f$ of the data, indicating an average 
	value for the fractal 
	dimension $d_f = 1.56 \pm 0.05$.$^{\ref{Tzsch94}}$
	}
\label{fig:nr}
\end{figure}

\begin{figure}[tb]
\centerline{
	\psfig{file=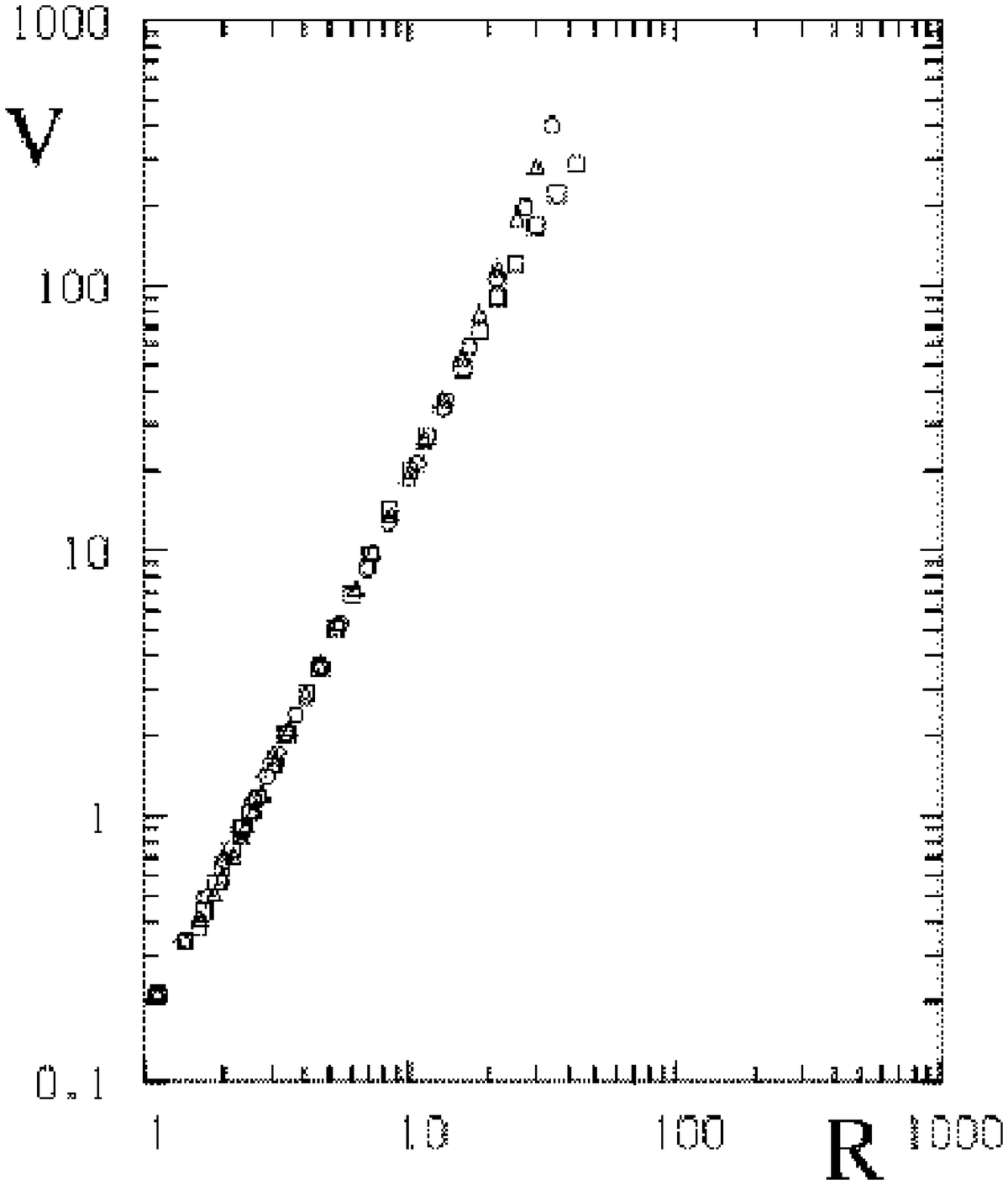,width=0.50\textwidth}
        }
\caption{
	Plot of the crack opening volume $V$ as a function of the
	average crack radius $R$ for fractal cracks with dimension 
	$d_f = 1.56$ using ($\triangle$) free boundary conditions and 
	($\Box$) periodic boundary conditions. For comparison we 
	show also results for a straight crack ($\circ$) 
	employing free boundary conditions. In all three cases 
	we find the behavior $V\sim R^2$ (in $d=2$).$^{\ref{Tzsch94}}$ 
}
\label{fig:vr}
\end{figure}
We briefly discussed in the introduction, Sec.~\ref{sec:intro}, 
Diffusion Limited Aggregation and its associated fractal aggregates.
Essentially DLA clusters growth by adding one particle per ``simulation
step'' to the clusters surface with a probability, $p_{ij}$, 
proportional to the local field gradient $\vert \nabla P\vert$ 
obtained from Laplace's equation, $\Delta\,P=0$. Similar 
considerations apply for the quasistatic growth of electrical 
discharge patterns, where an elementary discharge event is assumed 
to happen with a probability proportional to the local 
electrical field.$^{\ref{Witten}}$ 
Typically the boundary conditions 
are Dirichlet conditions for the external and the 
cluster surface. 

There have been attempts in literature 
to generalize the DLA model in several ways. A main 
proposition was the introduction of a growth exponent, $\eta$, 
describing how strong the growth probabilities are correlated to 
the gradients, $p_{ij}\sim \vert \nabla P\vert ^{\eta}$. 
For $\eta \to \infty$ the probability distribution 
becomes arbitrarily narrow (a delta peak) and the model is completely
deterministic, i.e., the cluster grows {\em always} at the 
surface location of highest field gradient (current). The resulting 
structures are usually linear in two dimensions. 
The case $\eta = 1$ corresponds to fractal DLA cluster growth, as already 
mentioned above. For $\eta = 0$ the cluster growth is completely
controlled by fluctuations and independent of the field solution. 
The corresponding structures are usually spatially compact.

It is important to note that the growth rule and in particular 
the employed growth exponent $\eta$ is completely independent of the 
investigated field equation. Rather the growth rule bases 
on plausible considerations.

There has been subsequently work been done to apply stochastic 
growth rules onto elastic systems, especially applying them to
crack growth problems. The ``Gradient'' in the growth rule is then  
typically replaced by the hydrostatic strain (eigenvalue) component.  
For certain classes of tensile problems fractal crack growth 
was found on two dimensional Hookean spring networks, exhibiting a fractal 
dimension $d_f \approx 1.6$ 
for $\eta =1$.$^{\ref{Louis},\ref{Meakin},\ref{p4}}$ 
This value for $d_f$ is relatively close to the one observed in a DLA
problem. 
However, the first author of this review conducted simulations 
on two dimensional beam lattices (see Sec.\ref{subsection:elastic_equations})
where a small preexisting crack was 
loaded at constant pressure from inside, and found non-fractal 
rather compact crack growth, already for $\eta =1$.$^{\ref{Tzsch94}}$ 
It was concluded that compressed microstructural elements
(beams) should not be considered for the breaking process because 
of the very asymmetry of interatomic potentials. 
With this modified breaking rule also non-fractal cracks were found 
for $\eta =1$. The cracks were, however, now linear in shape.
In order to obtain in a hydraulic fracture simulation
persisting branching cracks, much stronger fluctuations need to occur
than obtainable with $\eta =1$. This is not so surprising as almost
the whole elastic bulk is in a compressed state, while just around the
crack tips there exist very narrow cones of elements being in a stretched
state.
In Fig.~\ref{fig:Photoelast} we show the crack of a single, very
large simulation for the case of maximum
fluctuations, i.e., $\eta = 0$ using the above mentioned asymmetric
breaking criterion. Note that the breaking probability does {\em not}
depend on the strain magnitude, however, all compressed beams 
are assigned a breaking probability identically to zero.   
The hydraulic pressure inside the crack was kept constant and the
external boundaries are periodic in displacements, i.e., spanned on a torus.
Furthermore the system is assumed as impermeable and there are no
losses of the hydraulic fluid. 
The resulting crack structures are scaleinvariant (fractal) as 
can be seen in Fig.\ref{fig:nr}. The measured fractal dimension 
$d_f = 1.56\pm 0.05$ is slightly below the numerical value for tensile
problems, and slightly above the experimental value $d_f=1.41$
obtained from hydraulic fracturing of clay pastes in two dimensions, 
compare Fig.\ref{vandamme_fig}.

The total crack opening volume of the ramified cracks has been 
calculated and monitored during the growth, and plotted versa 
the average crack radius. This is shown in Fig.\ref{fig:vr}.
The crack opening volume for fractal cracks shows the same
dependency on (average) cracks size in comparison to 
straight cracks, namely $V\sim R^2$.    

We note that the foregoing considerations and results 
are for {\em impermeable} elastic materials, i.e., the influence 
of possible effects due to hydraulic fluid flow (hydraulic losses) 
have been neglected.

\section{Hydraulic Fracturing with structural disorder}
\label{sec:quenched}
The concept of a local cohesion strength has been used 
in a number of papers.$^{\ref{Sahim86}}$ 
One  assumes that a deformed elastic element 
connecting sites $i$ and $j$ breaks 
above a certain material specific threshold force 
$F^{ij}_{th}$ ('cohesion strength`). This kind of disorder is 
also called deterministic or quenched disorder in the literature,
because once the breaking thresholds have been chosen the model 
behaves in a completely deterministic way, contrary to the model
described in the foregoing section. However, we prefer the term 
`structural disorder' as a spatial variation in material properties
is frequently associated with certain micro-structural
particularities within a material.       
If the inner stresses are 
above their thresholds the beams are broken and removed, i.e. their 
elastic moduli are set to zero.
Since the cohesional strength for 
compression is much higher than for tension we assume 
that compressed beams cannot break.$^{\ref{Tzsch94}}$  
The breaking thresholds are initially
distributed randomly according to some probability density function,
$\rho (F_{th})$. 
At this point it would be ideal to employ empirical strength
distributions of the considered material for $\rho (F_{th})$.
However, at the present stage of research we are basically interested 
in generic features of such crack growth models.
The width of the strength distribution is most easily controlled 
employing power-law distributions, 
\begin{equation}
\rho (F_{th})\sim F_{th}^r,
\label{equation:disorder}
\end{equation} 
with
$F_{th}\in [0,F_{max}]$ and $r > -1$. 
Negative exponents $r$ are used to describe strong
cohesive disorder (broad distributions)
while large positive exponents 
correspond to weak disorder (narrow distributions).
It is convenient to express 
the normalization factor and $F_{max}$ by the distribution's expectation
value $\langle F_{th} \rangle$ and the exponent $r$. This allows 
to investigate disorder effects due to deviations around 
$\langle F_{th} \rangle$ for constant $\langle F_{th} \rangle$.

In the following we will discuss the modelling of crack growth in 
impermeable and permeable materials separately
because the employed breaking criteria slightly
differ for both cases. 

\subsection{The impermeable medium}
\label{subsec:impermeable}

\begin{figure}[tb]
\centerline{
	\psfig{file=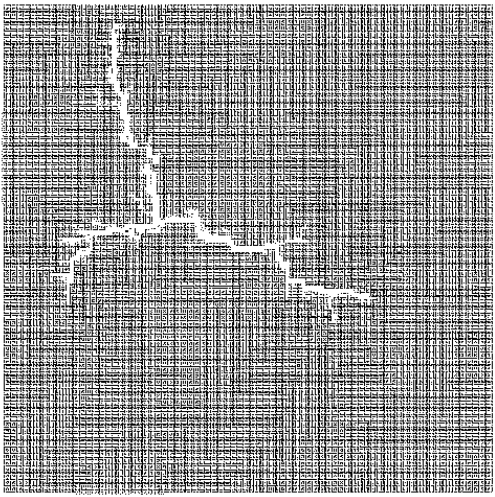,width=0.50\textwidth}
        }
\caption{
	Typical hydraulic crack for strong disorder,
        $r=-0.7$ and $\langle F_{th}\rangle  =0.01$, on a lattice of 
        size $L=150$.  
        The crack consists of $629$ broken beams after $1500$
        time steps using a hydraulic flux $q_{in} =0.05$.
        The injection point is the center of the lattice. 
        We have used periodic boundary 
        conditions in vertical and horizontal directions.$^{\ref{Tzsch95}}$
	}
\label{fig:95precrack}
\end{figure}

\begin{figure}[tb]
\centerline{
	\psfig{file=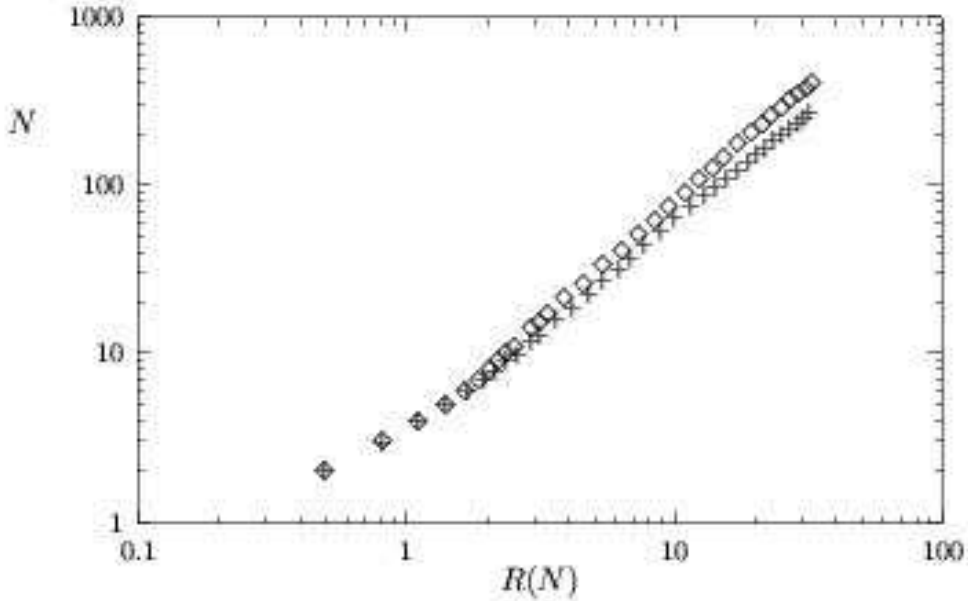,width=1.00\textwidth}
        }
\caption{
	Log-log plot of the number $N$ of broken beams 
        versus the average radius $R(N)$ for two different 
        distributions of breaking thresholds: ($\diamond$) $r=-0.7$, 
        averages over $60$ cracks,  fractal 
        dimension $d_f =1.44\pm 0.10$; ($+$) $r=-0.5$, 
        averages over $53$ cracks, fractal dimension $d_f=1.39\pm 0.10$. 
        For all simulations we have used an average cohesion value 
        $\langle F_{th} \rangle =0.01 $ and a  
        linear lattice size $L=150$.$^{\ref{Tzsch95}}$
	}
\label{fig:95predf}
\end{figure}

\begin{figure}[tb]
\centerline{
	\psfig{file=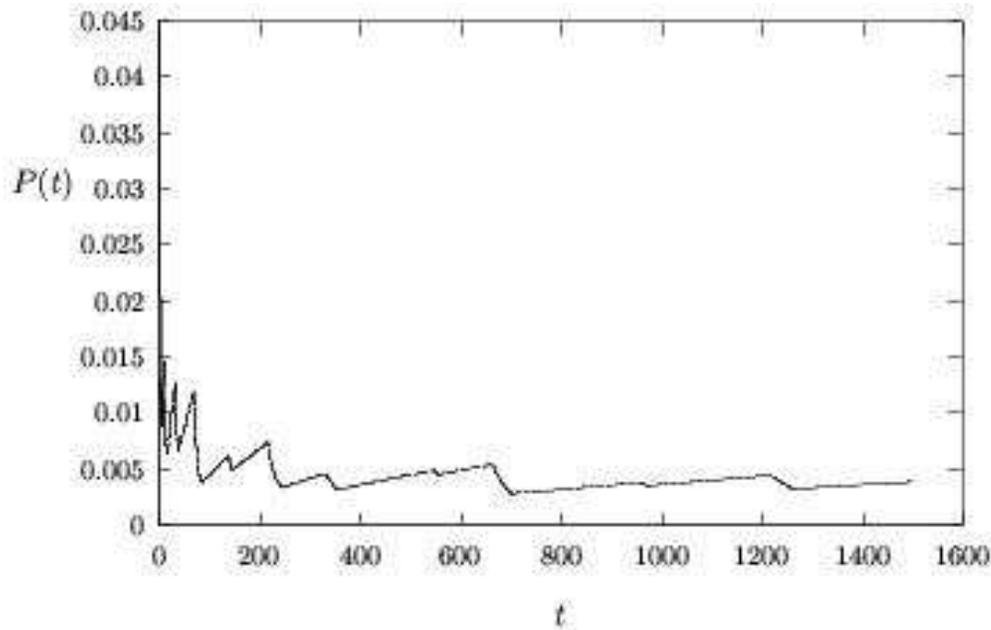,width=1.00\textwidth}
        }
\caption{
	Linear plot of the pressure $P$ inside the crack versus time 
	$t$. The record corresponds to the simulation displayed 
	in Fig.~\ref{fig:95precrack}.$^{\ref{Tzsch95}}$
	}
\label{fig:press_erratic}
\end{figure}

\begin{figure}[tb]
\centerline{
	\psfig{file=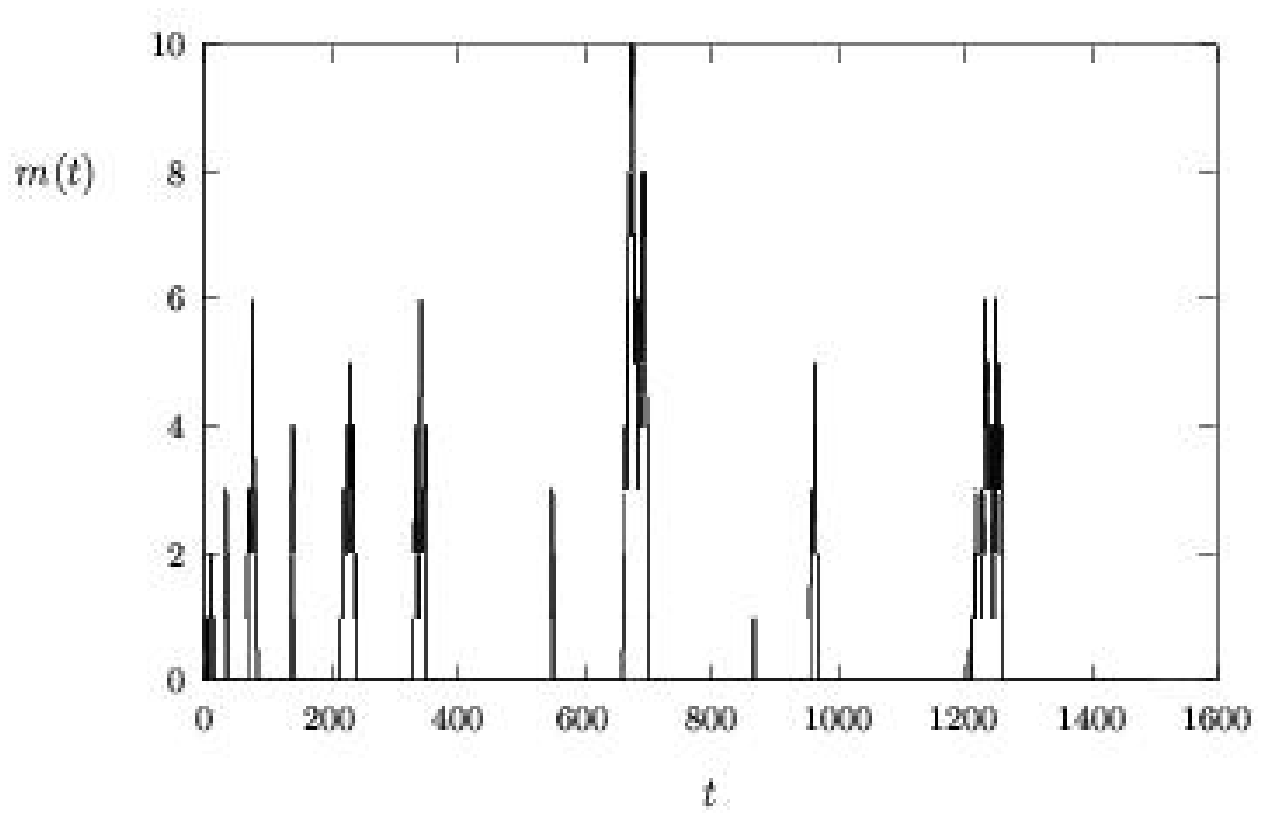,width=1.00\textwidth}
        }
\caption{
	Temporal breaking sequence corresponding to the 
	crack displayed in Fig.~\ref{fig:95precrack}. 
	The magnitude $m(t)$ is 
	defined as the number of per unit time step simultaneously
	broken beams. The simulation was stopped after $1500$ time 
	steps. Note the clustering of breaking events and the large
	time intervals of quiescence.$^{\ref{Tzsch95}}$
	}
\label{fig:bursts}
\end{figure}

The hydraulic fracturing of an impermeable solid represents 
a limiting case due to the vanishing hydraulic losses. For 
a porous material saturated by an incompressible fluid this 
would imply that {\em only} pure shear deformations would be possible.
This follows directly from the time-dependent flow equation, i.e., 
for $k\to 0$ or $\mu \to \infty$ Eq.~(\ref{equation:pressure}) reduces
to the condition $\Div {\bf u}= \mbox{const.}$ preserving the volume 
of a deformation. This is not a very typical experimental situation.
However, if the pores are filled by very compressible gases one might 
even neglect the deformation dependent volume forces for impermeable
solids.

In the following we consider the case where an incompressible fluid
hydraulically propagates a crack through an impermeable material.
We require that the pores are filled by a very compressible
substance compared to the solids compressibility. Such situation 
is approximately encountered for magma driven 
cracks.$^{\ref{Spence85}}$
    
We consider a constant fluid injection rate, 
$q_{in}=5\cdot 10^{-2}l^2d$, into the 
hydraulic crack and a vanishing pore pressure, $p_0 = 0$.
The boundary conditions have been discussed 
in Sec.~\ref{subsection:boundary_conditions}, i.e., see 
Eq.~(\ref{equation:lambda_equation_discrete}) for vanishing Darcy 
losses, $D^* = 0$.
The simulation procedure follows the mentioned steps \ref{a}-\ref{i}
in Sec.~\ref{subsection:boundary_conditions} with the exception 
that in the present case no pressure equations need to be solved.
Instead the `test pressure' $p^*$ does appear directly as 
inhomogeneity for the elastic equations along the crack contour.
Initially one small crack (one broken vertical beam) is prepared on the
grid center.  
For the distribution of breaking thresholds, 
Eq.~(\ref{equation:disorder}), the values 
$\langle F_{th}\rangle = 10^{-2}$ (arbitrary units), 
$r =-0.5$ and  $r=-0.7$ (strong disorder), and $r=+\infty$ (no
disorder) are used.
After each time increment, $\Delta\,t = 1$, {\em all} beams 
carrying {\em tensile} forces larger than the associated breaking
thresholds are broken, i.e., their elastic moduli are 
set to zero.

For $r=+\infty$ (no disorder) straight cracks are observed and the
hydraulic crack pressure drops in form of a power-law in time,
$P(t)\sim t^{-1/3}$.$^{\ref{Tzsch95}}$ 

For strong disorder the situation becomes more complicated.
In Fig.~\ref{fig:95precrack} we show a calculated sample crack. The
obtained crack is a branching structure exhibiting many small scale
branches but only a very few branches on the larger length scale.
A comparison with the crack structure shown in
Fig.~\ref{fig:Photoelast} reveals a much lower crack density in the
present case. This is directly measured in Fig.~\ref{fig:95predf} by
plotting the number of broken beams $N$ in dependence of the average 
crack size $R$,$^{\ref{radius}}$  
yielding the cracks fractal dimension $d_f$ using 
the relationship $N\sim R^{d_f}$. The fractal
dimensions for the investigated threshold distributions 
$r=-0.7$ $(\diamond)$ and $r=-0.5$ $(+)$ take the values 
$d_f=1.44\pm 0.10  $ and $d_f=1.39\pm 0.10 $ respectively.
These values for $d_f$ 
are consistent with the fractal dimension found in
the two dimensional experiments on hydraulic fracturing of
viscoelastic clays.$^{\ref{vandam}}$

The time record of the hydraulic pressure at the injection point is
of considerable interest because on the one hand it is often experimentally 
measurable and on the other hand does it contain information about 
rheological and breaking properties.
If the considered crack is at rest the hydraulic pressure $P$ will 
generally build up in time (loading). For an impermeable medium 
the pressure build up will correspond to the materials elastic
response on the injected fluid (and possible fluid flow effects 
within the crack which are neglect in the present case).
We consider the fluid as incompressible implying a linear relationship
between crack volume and time, and due to the linearity and time 
independence of employed elastic equations a linear increase of 
pressure in time is the consequence. The associated slope is related to the 
materials mechanical stiffness (for the present crack).
However, if the crack starts to move the stiffness decreases and so does
the hydraulic pressure. There exists a cutoff for this pressure 
drop because the materials average cohesion is non-zero. 
Therefore the crack comes to rest again after some time period.  
Figure \ref{fig:press_erratic} displays such an `erratic' pressure 
record. The strongest pressure fluctuations occur for small cracks
(early times) at high frequency whereas lateron the loading periods
become larger and larger. This is due to the rapidly decreasing 
stiffness as soon as the crack advances.
The loading periods are periods of `quiescence' (no breaking); 
the unloading periods correspond to `bursts' (temporally localized 
groups of breaking events).
This is exemplified in Fig.~\ref{fig:bursts}.
The temporal correlations of such bursts have been analyzed in detail in 
Ref.~\ref{Tzsch95}.    

\subsection{The permeable medium}
\label{subsec:permeable}

This section considers hydro-fracturing of a permeable porous medium,
where an incompressible fluid is injected at the center of a quadratic grid
at a constant rate, $q_{in}$.  
The boundary conditions have been discussed in 
Sec.~\ref{subsection:boundary_conditions}.

The dimensionless formulation of the pressure equation is first
explained before we show some examples.  The considered pressure equation
Eq.~(\ref{equation:pressure_discrete}) is made dimensionless by considering
the forces acting on the beams surrounding the injection cell at
the center of the grid.  Darcy's law yields the Darcy velocity
$v=-(k/\mu)dp/dx$ in the $x$-direction.  The Darcy velocity is also
$v$ in the $y$-direction when assuming symmetry.  The forces acting
on the beams surrounding the injection cell are $F=l^2d\,dp/dx$,
because the pressure difference between two neighbor cells is
$dp=(dp/dx)l$.  The length of a beam is $l$ and the width of a beam
is $d$.  The beam force $F$ can be written $F=\mu vl^2d/k$, when
Darcy's law is used to replace the pressure gradient $dp/dx$ with
the Darcy velocity $v$.  The volume flux into the center cell is
initially $q_{in}=vl^2$.  The force $F$ is written in terms of $q_{in}$ rather
than $v$, and we have $F=\mu q_{in} d /k$.  This force is scaled with
the unit force $F_0=Ed^2$, (where $E$ is Young's modulus), which
can be written $F/F_0=q_{in}/q_0$.  The unit injection rate becomes
$q_0=kEd/\mu$.

This scaling shows two things.  Firstly, it shows that a dimensionless 
injection rate of order one or larger implies strong enough forces on
the center cell for deformations on grid to be noticeable.  Secondly,
the (dimensionless) average beam strength must be less than $q_{in}/q_0$ for
a crack to nucleate.

The characteristic time becomes $t_0=\mu l^2/(Ek)$, which will
be as low as nano seconds for a grid with beam length $l=10^{-5}$~m,
permeability $k\sim 10^{-15}$~m$^2$, Young's modulus 
$E\sim 10^{11}$~Pa, and fluid viscosity $\mu \sim 10^{-3}$~Pa\,s.
We consider a constant permeability throughout the material.

The full pressure equation for an incompressible fluid
Eq.~(\ref{equation:pressure}) is not solved.  The pressure
equation is therefore not fully coupled to the volume changes in
both the bulk and the fracture.  However, it is still possible to
couple the hydraulic pressure to the elastic deformations of the crack by
considering conservation of fluid, as shown in
Sec.~\ref{subsection:boundary_conditions}.  The (Laplacian) pressure
equation is first solved for a unit pressure $p^*$ on the crack contour.
The total Darcy flux $D^*$ is determined from
this solution; the corresponding crack volume $V^*$ is computed from
the associated elastic solution.
The scaling parameter $\lambda\,(t)$
yielding the physical pressures and strains is computed 
each time step according Eq.~(\ref{equation:lambda_equation_discrete}),
see Sec.~\ref{subsection:boundary_conditions} steps \ref{a}-\ref{i}.  

The beams on the crack surface are checked after each time
step to see if any of them are stretched beyond their strength.
If there are any over-stretched beams, then the most over-stretched
beam will be broken.  
We showed in the foregoing section that the advance of fracture is
localized in time in form of bursts, separated by large time periods 
of quiescence (loading).
The time-step used during the bursts
is therefore ``small''.  On the other hand, the quiet periods
between two bursts can be very long compared to the short time span of
a burst.  The time-step is therefore multiplied by a factor 
larger than $1$ after time-steps where no beams were broken.  This 
factor could be chosen so that ten such steps together yield a
combined factor equal to for instance $2$ or $10$.  Increasing
each time-step between the bursts by a factor makes it possible
to obtain a step length sufficiently long to get through these
relatively long periods of pressure build-up.  However, the
time-step is immediately set to its minimum value once a new
beam breaks.
 
Three numerical examples are studied, which differ in the average 
beam strength $\langle F_{th} \rangle$. 
The average beam strength in these cases and the 
injection rate $q_{in}$ are given in the Table~\ref{tab:cases}.
\begin{table}\begin{center}\begin{tabular}{ccc}\hline\hline
Case       & Average beam strength & Injection rate   \\
No         & $\langle F_{th}\rangle/F_0$    & $q_{in}/q_0$          \\
\hline
I          & $10^{-2}$             & $10^{-3}$        \\
II         & $10^{-3}$             & $10^{-3}$        \\
III        & $10^{-4}$             & $10^{-3}$        \\
\hline\hline
\end{tabular}\caption{The numerical cases have the same 
injection rate $q_{in}$, but differ for the average beam 
strength $\langle F_{th} \rangle$.}
\label{tab:cases}\end{center}\end{table}
All three cases have the grid size $100\times 100$ and a beam
strength distribution characterized by the exponent
$r=-0.9$, compare Eq.~({\ref{equation:disorder}}).
  
A value for $r$ close
to $-1$ indicates that the beams have very broadly distributed
strengths.  The employed material and geometry dependent constants are 
$a=1$, $b=2.5$ and $c=1200$, when expressed dimensionless.
(The scaling of these constants is such that $a$ becomes $1$.) 

\begin{figure}[tb]
    \centerline{\hbox{
    \psfig{file=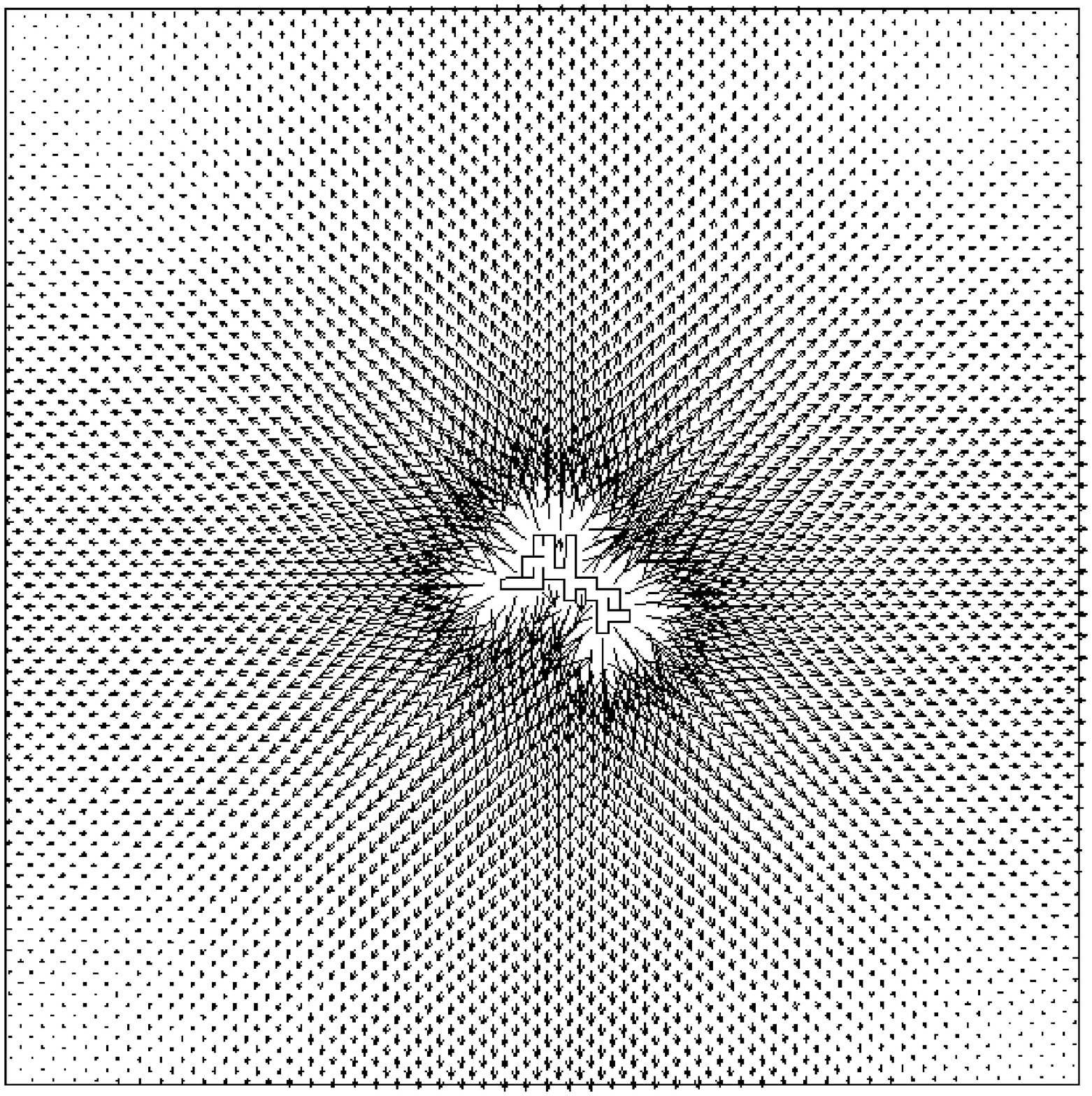,width=0.5\textwidth}
    \psfig{file=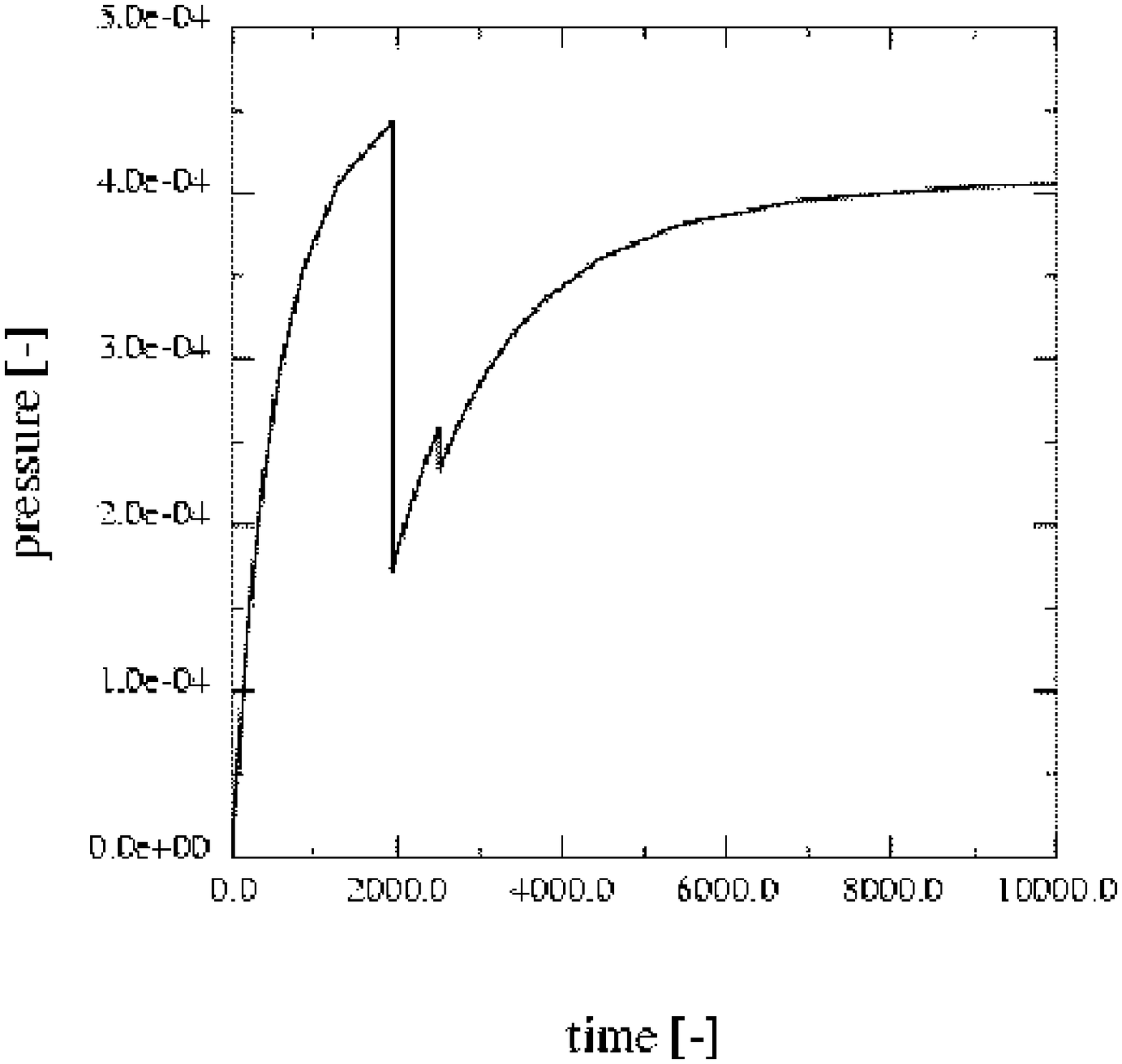,width=0.5\textwidth}
    }}
    \caption{Case~I has an average beam 
	strength~$\langle F_{th}\rangle/F_0=10^{-2}$ 
	and an
        injection rate $q_{in}/q_0=10^{-3}$.  
        (a) (left) The final state of the branching crack structure and
        the corresponding Darcy flow field.  (b) (right) 
        The hydraulic pressure in the crack as a function 
	of time.$^{\ref{Wangen}}$
	}
    \label{fig:case1}
\end{figure}
\begin{figure}[tb]
    \centerline{\hbox{
    \psfig{file=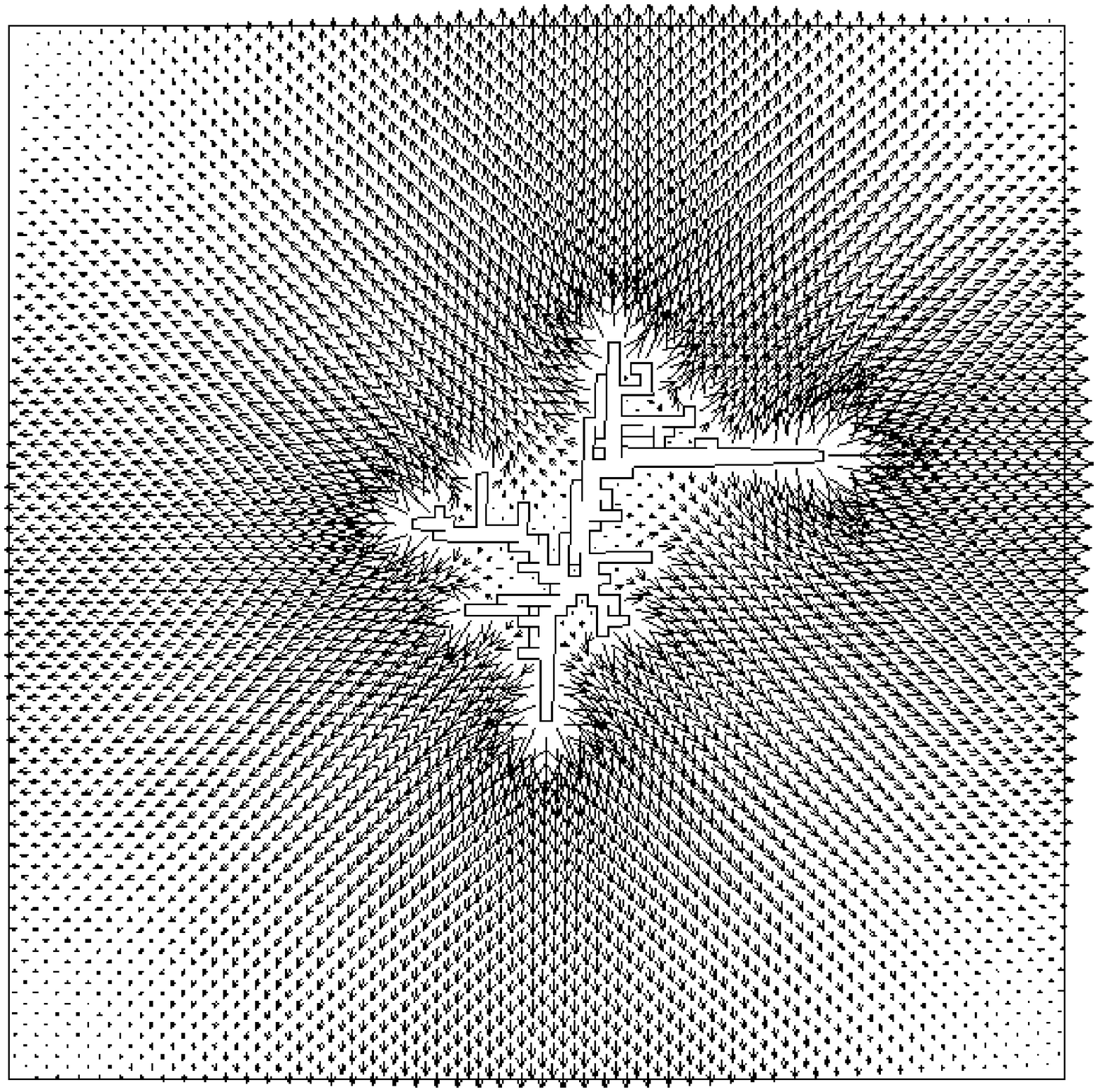,width=0.5\textwidth}
    \psfig{file=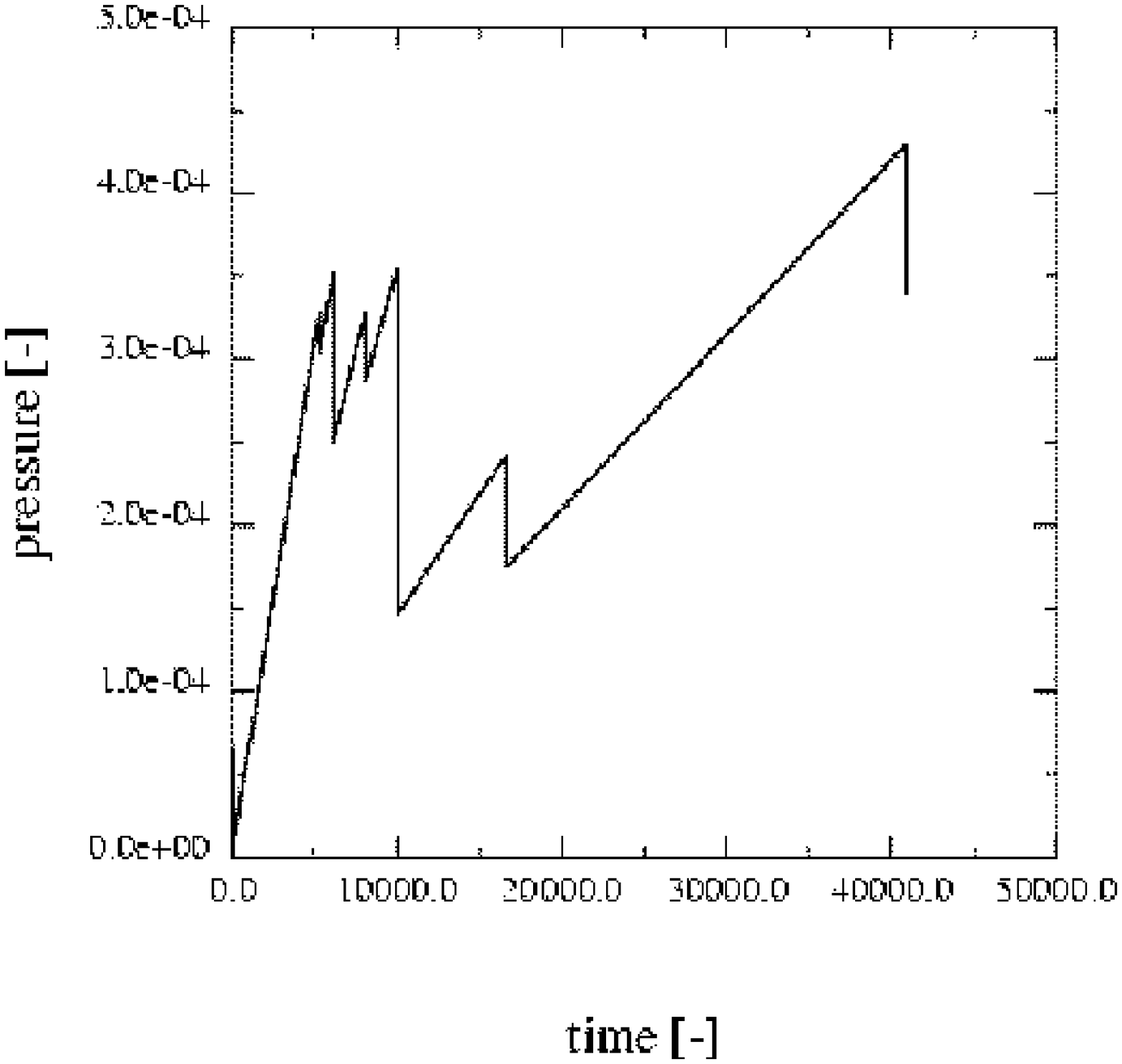,width=0.5\textwidth}
    }}
    \caption{Case~II has an average beam 
	strength~$\langle F_{th}\rangle/F_0=10^{-3}$
    	and an
        injection rate $q_{in}/q_0=10^{-3}$.  (a) (left) The crack and the 
        corresponding Darcy flow field. (b) (right) The hydraulic pressure 
        in the crack as a function of time.$^{\ref{Wangen}}$
	}
    \label{fig:case2}
\end{figure}
\begin{figure}[tb]
    \centerline{\hbox{
    \psfig{file=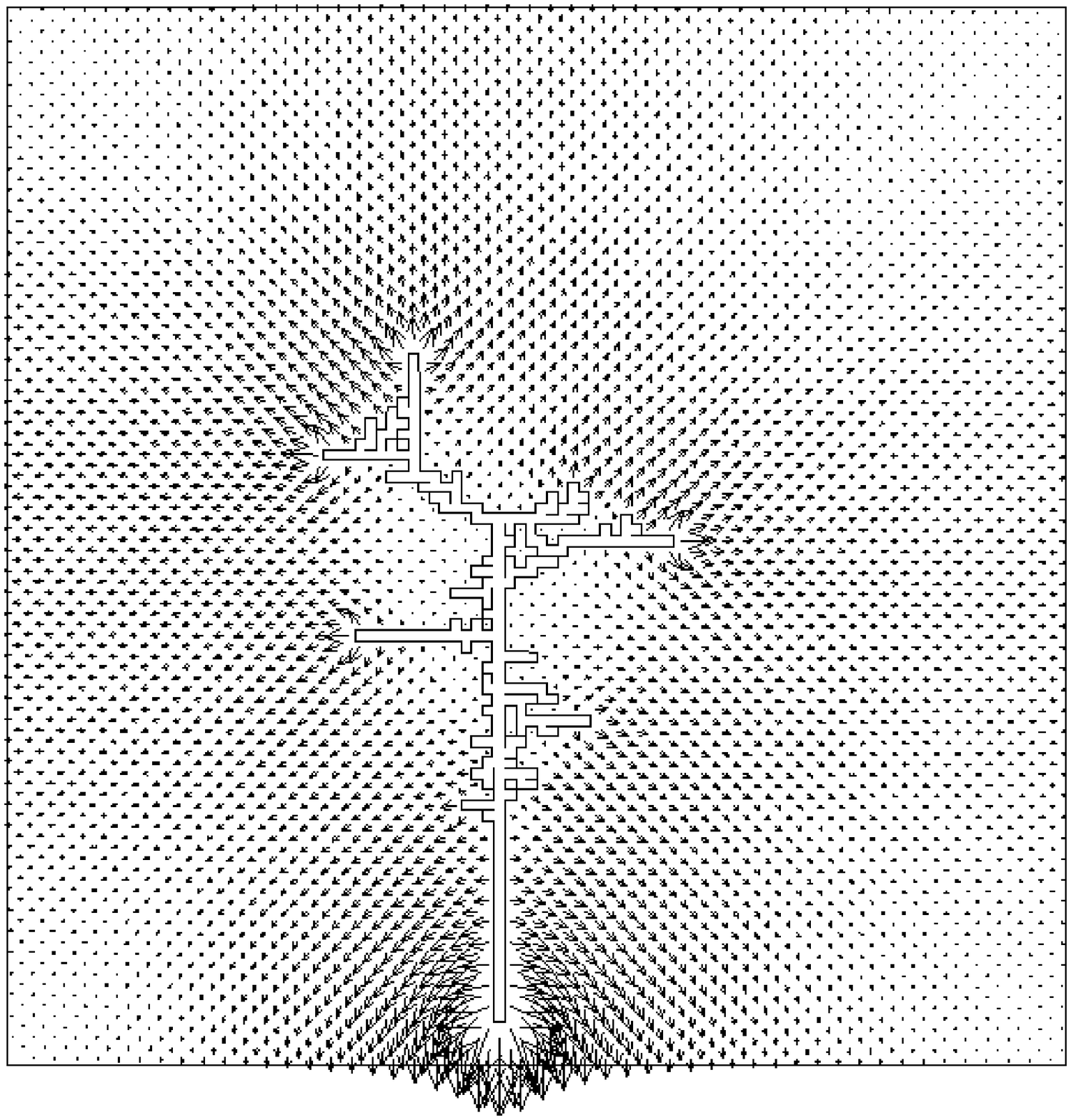,width=0.5\textwidth}
    \psfig{file=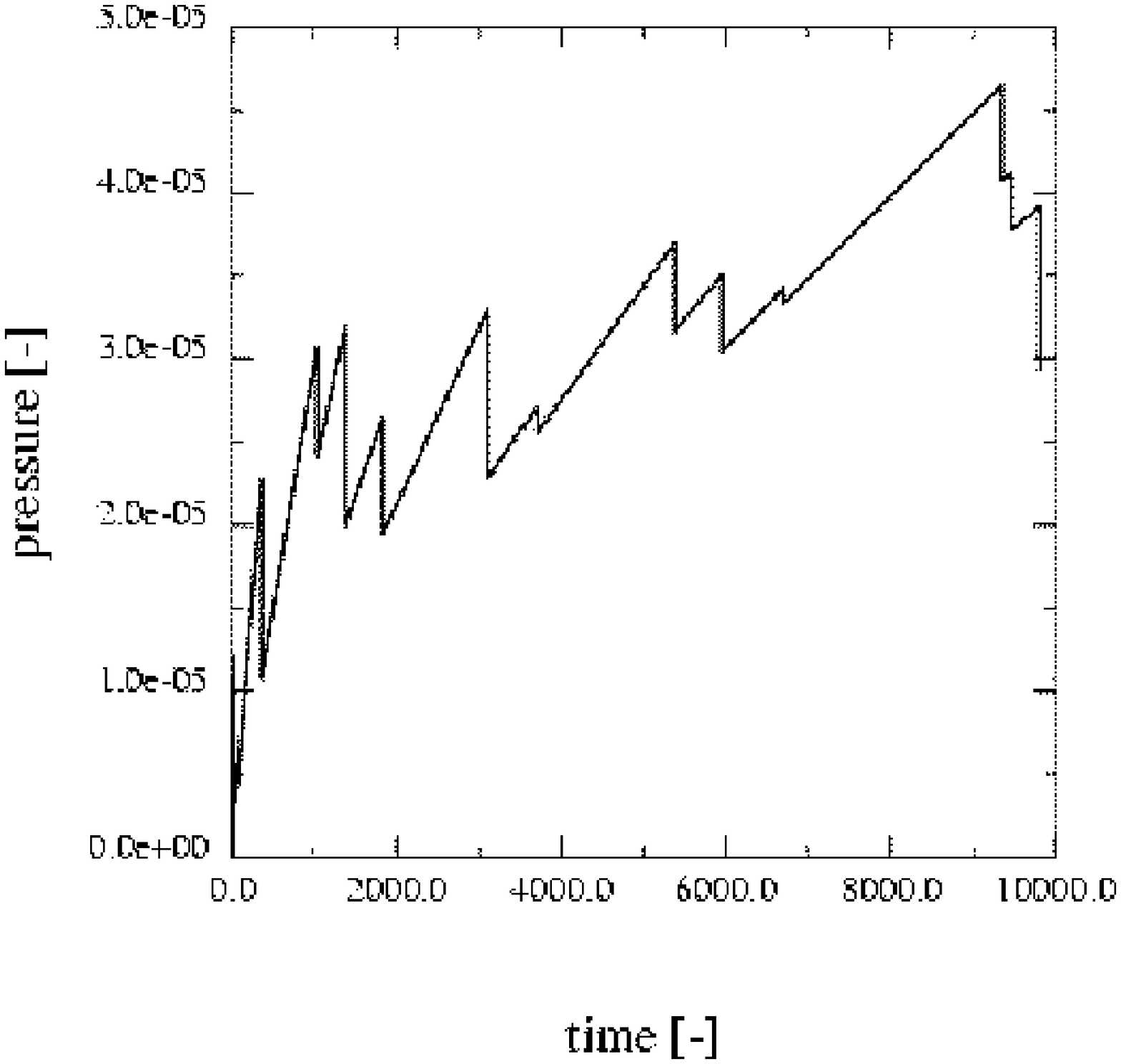,width=0.5\textwidth}
    }}
    \caption{Case~III has an average beam 
	strength~$\langle F_{th}\rangle /F_0=10^{-4}$ 
	and an
        injection rate $q_{in}/q_0=10^{-3}$. (a) (left) The crack and the 
        corresponding flow field. Note the screening of the flow field
	within the crack branches.  
	(b) (right)
        The hydraulic pressure in the crack as a function 
	of time.$^{\ref{Wangen}}$
	}
    \label{fig:case3}
\end{figure}

\begin{figure}[tb]
    \centerline{\hbox{
    \psfig{file=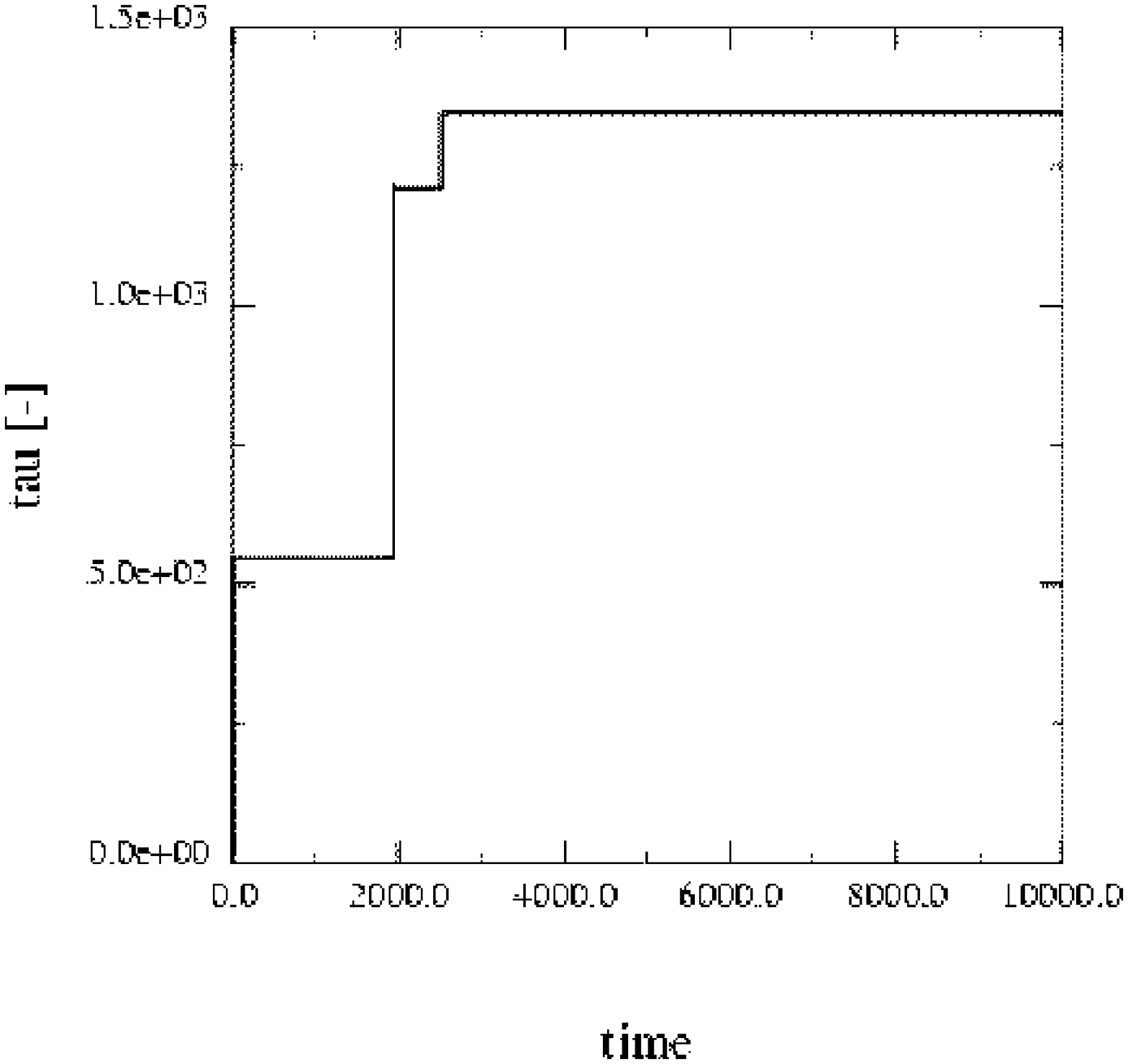,width=0.35\textwidth}
    \psfig{file=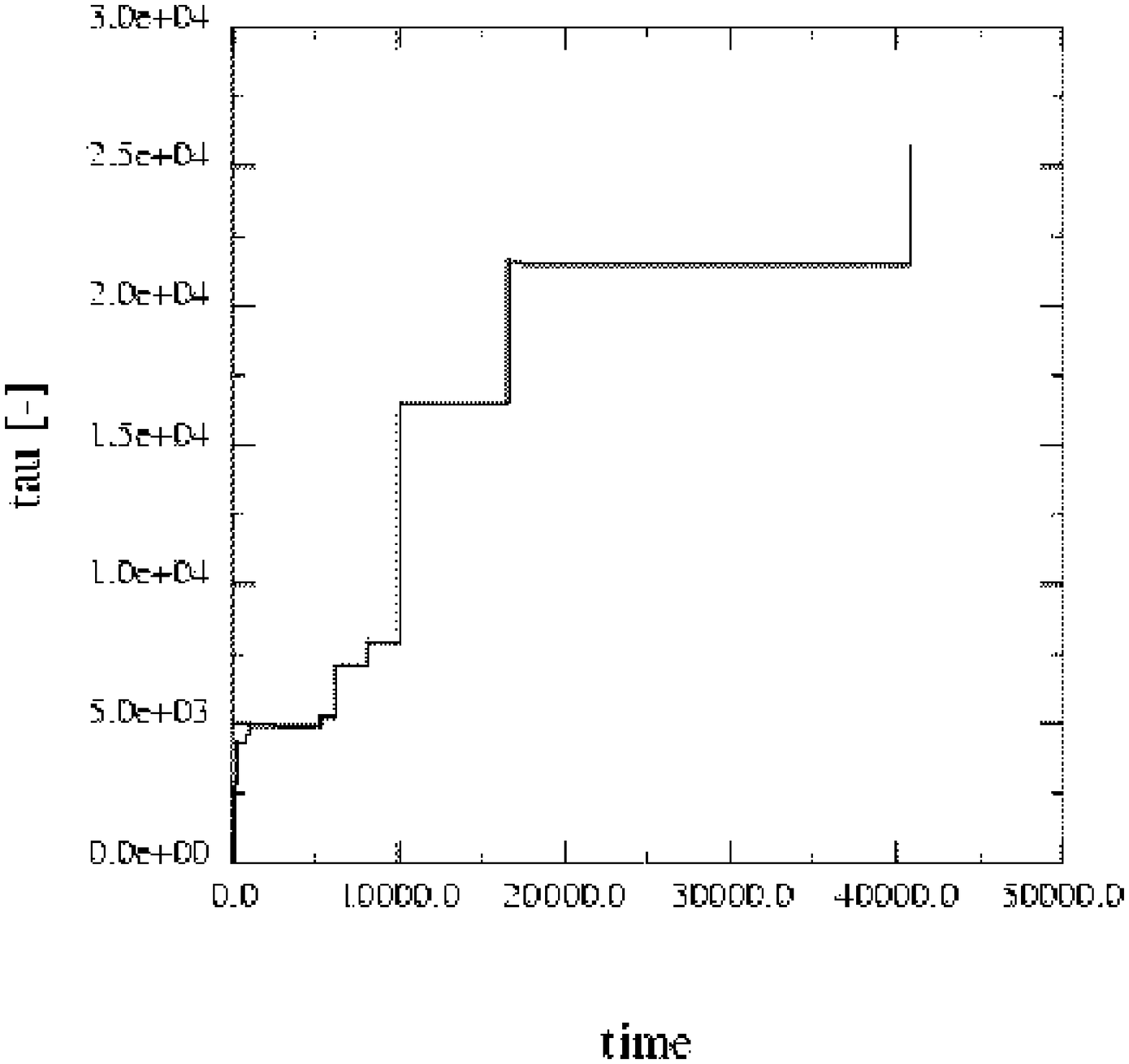,width=0.35\textwidth}
    \psfig{file=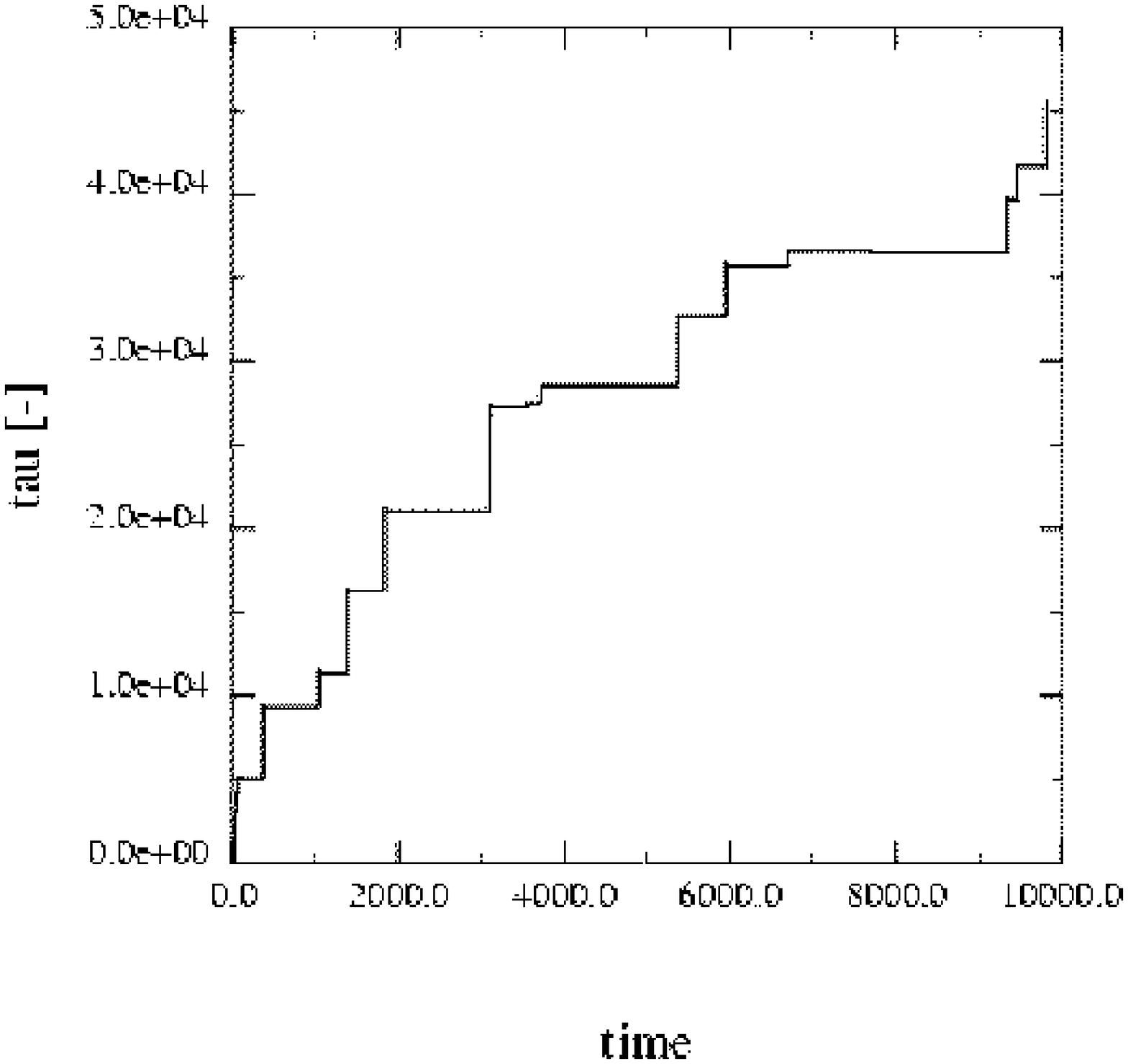,width=0.35\textwidth}
    }}
    \caption{The characteristic time 
	$\tau = \int {\bf u^*}\cdot d\,{\bf A} /
	\int {\bf v^*}\cdot d\,{\bf A}$ 
	as a function of crack growth for the
        three cases of Table~\ref{tab:cases}.}
    \label{fig:tau}
\end{figure}

The crack in case~I, where 
$q_{in}/q_0\ll \langle F_{th}\rangle /F_0$, is shown in
Fig.~\ref{fig:case1}a.  This figure shows the crack after it
stopped growing.  The hydraulic pressure is shown in
Fig.~\ref{fig:case1}b, where it is seen that the pressure reaches
a stationary value.  A small crack nucleated in this
case, even though the average beam strength is $10$ times larger
than the injection rate.  This is due to the strong heterogeneities
($r=-0.9$).  The pressure follows an exponential rise in
time according to the expression for the scaling parameter $\lambda$
in Eq.~(\ref{equation:lambda_solution}).  The first smooth
rise of the pressure ends by breaking beams until
no more beams are overstressed, 
and a second rise of pressure follows.  
This second rise leads to a stationary pressure.  Notice
that the first rise of pressure points towards a higher stationary
pressure than the second rise.  This is expected because the
maximum possible pressure in a constant crack geometry, 
given a constant injection rate, 
is decreasing with increasing crack size.
A large crack
requires a lower stationary hydraulic pressure compared to a small
crack in order to balance the injection rate to the hydraulic losses.  
The small crack seen in Fig.~\ref{fig:case1}a
is almost entirely due to the first burst.  After
the first burst, the pressure cannot reach a sufficiently
high level any more to break more beams.  
The time $\tau$ (see Eq.~(\ref{equation:lambda_constants})) 
is plotted in Fig.~\ref{fig:tau}a.   
Because $\tau$ measures the typical elastic response versus the 
typical flow response, it defines the characteristic time scale
below which the (re)loading process appears to happen within an 
impermeable medium. 
It is seen that $\tau$ is
increasing in crack size, which is consistent with the 
pressure plot Fig.~\ref{fig:case1}b.  
The characteristic time increases
with the crack size even though the stationary pressure is decreasing.

The case~II shows crack propagation for an average
beam strength equal to the injection rate, see Table~\ref{tab:cases}.
The forces from the fluid gradients on the beam structure are
then comparable with the strength of the beams.  The pressure during
the cracking process is shown in Fig.~\ref{fig:case2}b, and it is
seen that the pressure is rising almost linearly between the bursts
of beam breaking.  The small initial bursts were sufficient to
generate a crack with a time constant $\tau$ much larger than
the time span shown in the pressure plot.  This explains why the
pressure rise appears almost linear rather than exponential in time 
as given by Eq.~(\ref{equation:lambda_solution}).
The characteristic time $\tau$ is shown in
Fig.~\ref{fig:tau}b.  The crack structure at the end of the time
period in the pressure plot (Fig.~\ref{fig:case2}b) is shown in
Fig.~\ref{fig:case2}a.  The crack in this case of broadly distributed
beam strengths shows the growth of a branching crack structure.
Note that within the crack `fjords' the Darcy flow almost vanishes.

Case~III has an average beam strengths $\langle F_{th}\rangle / F_0$
one order of magnitude lower than the injection rate $q_{in}/q_0$.  
The injection rate is therefore high compared to strength
of the structure.  This is also seen from the pressure plot in
Fig.~\ref{fig:case3}b, which is characterized by more frequent
bursts compared to case~II.  Only a slight pressure
increase is enough to trigger a new burst.  The hydraulic pressure
level of beam breaking in Fig.~\ref{fig:case3}b is about one order
of magnitude lower than the corresponding pressure level in
figure~Fig.~\ref{fig:case2}b.  This is as expected from the average
beam strengths of these cases.  The characteristic time $\tau$ for case~III
is shown in Fig.~~\ref{fig:tau}c, which also shows a large number
of bursts.  The crack in its final state is shown 
in Fig.~\ref{fig:case3}a.  
A branching crack structure is seen, where the branch 
close to the boundary seems to be growing faster than the other branches.  
This seems to be a boundary effect because the pressure
gradients close to a boundary are larger. 
This effect favourizes growth of cracks close to the
boundaries.

The presented cases demonstrate several features of hydraulic fracturing
in a permeable medium with broadly distributed cohesion, when
a fluid is injected at a constant rate at the center of the test
sample.  They clearly show how the hydraulic pressure is increasing during
``quiet'' periods.  The pressure increases according
to Eq.~(\ref{equation:lambda_solution}). 
The characteristic time $\tau$ in 
Eq.~(\ref{equation:lambda_constants}) is
computed, and it is seen that $\tau$ becomes large compared to
the time period between successive bursts.  The pressure
therefore appears to rise linearly between the bursts.  The 
characteristic time
$\tau$ is also increasing with the crack size, which
implies that the time periods between successive bursts get longer as
the crack growths.  This can also be seen from the slopes in
Figs.\ref{fig:case1}b-\ref{fig:case3}b
which are decreasing with increasing crack size.
The pressure drops during the bursts. 
Furthermore, the bursts also become more frequent when the
average beam strength $\langle F_{th}\rangle /F_0$
gets low compared to the injection rate $q_{in}/q_0$.
\section*{Conclusion}
We have reviewed some numerical results for hydraulic fracturing 
in porous materials employing microstructural fracture growth models.  
The considered models are {\em not} finite difference schemes of 
certain continuum mechanical equations. Rather they implement an 
approach to crack growth problems in the spirit of Monte Carlo methods
of statistical mechanics. 
This allows to track quasistatic crack growth under very 
general conditions, i.e., for disordered systems on mesoscopic 
length scales.

\section*{Acknowledgments}
F.~T.~ would like to acknowledge financial support from
CEC under grant number ERBFMBICT 950009.

%
%


\begin{thebibliography}{10}
\bibitem{dual}\label{dual}
The dual lattice of a square lattice is also a square lattice,
however, displaced half a lattice constant both in $x$ and $y$ 
directions ($d=2$).
\bibitem{radius}\label{radius}
The typical radius $R$ of a branching crack structure is preferably
calculated as a radius of gyration, 
$R^2(N)={1\over N}\sum_i (\vec{r}_i - \vec{r}_0 
)^2$ with $\vec{r}_0 = {1\over N}\sum_i \vec{r}_i$.
\end{thebibliography}
\end{document}